\documentclass[sigconf]{acmart}
\usepackage[T1]{fontenc}
\usepackage{graphicx}
\usepackage{caption}
\usepackage{subcaption}

\graphicspath{{./figures/}}
\usepackage{multirow}
\usepackage{makecell}
\usepackage{enumitem}
\usepackage{cleveref}
\usepackage{hyperref}
\hypersetup{
    colorlinks=true,
    linkcolor=blue,
    filecolor=magenta,      
    urlcolor=cyan,
}

\usepackage[colorinlistoftodos,prependcaption,textsize=tiny]{todonotes}
\copyrightyear{2020}
\acmYear{2020}
\setcopyright{acmcopyright}\acmConference[CIKM '20]{Proceedings of the 29th ACM International Conference on Information and Knowledge Management}{October 19--23, 2020}{Virtual Event, Ireland}
\acmBooktitle{Proceedings of the 29th ACM International Conference on Information and Knowledge Management (CIKM '20), October 19--23, 2020, Virtual Event, Ireland}
\acmPrice{15.00}
\acmDOI{10.1145/3340531.3411934}
\acmISBN{978-1-4503-6859-9/20/10}
\begin{document}
\fancyhead{}
\def\BibTeX{{\rm B\kern-.05em{\sc i\kern-.025em b}\kern-.08emT\kern-.1667em\lower.7ex\hbox{E}\kern-.125emX}}
    
% Rights management information. 
% This information is sent to you when you complete the rights form.
% These commands have SAMPLE values in them; it is your responsibility as an author to replace
% the commands and values with those provided to you when you complete the rights form.
%
% These commands are for a PROCEEDINGS abstract or paper.

%
% The "title" command has an optional parameter, allowing the author to define a "short title" to be used in page headers.
\title[Unsupervised Cyberbullying Detection via Time-Informed Gaussian Mixture Model]{Unsupervised Cyberbullying Detection \\ via Time-Informed Gaussian Mixture Model}

\author{Lu Cheng$^\dagger$, Kai Shu$^\ddagger$, Siqi Wu$^\star$, Yasin N. Silva$^\dagger$, Deborah L. Hall$^\dagger$, Huan Liu$^\dagger$}
\affiliation{%
  \institution{$^\dagger$Arizona State University, $^\ddagger$Illinois Institute of Technology, $^\star$Australian National University}
}
\email{{lcheng35,ysilva,d.hall,huanliu}@asu.edu, kshu@iit.edu, siqi.wu@anu.edu.au}

%
% By default, the full list of authors will be used in the page headers. Often, this list is too long, and will overlap
% other information printed in the page headers. This command allows the author to define a more concise list
% of authors' names for this purpose.
\renewcommand{\shortauthors}{Cheng, et al.}
\fancyhead{}

\begin{abstract}
Social media is a vital means for information-sharing due to its easy access, low cost, and fast dissemination characteristics. However, increases in social media usage have corresponded with a rise in the prevalence of cyberbullying. Most existing cyberbullying detection methods are \textit{supervised} and, thus, have two key drawbacks: (1) The data labeling process is often time-consuming and labor-intensive; (2) Current labeling guidelines may not be generalized to future instances because of different language usage and evolving social networks. To address these limitations, this work introduces a principled approach for \textit{unsupervised} cyberbullying detection. The proposed model consists of two main components: (1) A \textit{representation learning} network that encodes the social media session by exploiting multi-modal features, e.g., text, network, and time. (2) A \textit{multi-task learning} network that simultaneously fits the comment inter-arrival times and estimates the bullying likelihood based on a Gaussian Mixture Model. The proposed model jointly optimizes the parameters of both components to overcome the shortcomings of decoupled training. Our core contribution is an unsupervised cyberbullying detection model that not only experimentally outperforms the state-of-the-art unsupervised models, but also achieves competitive performance compared to supervised models.
\end{abstract}

% Keywords. The author(s) should pick words that accurately describe the work being
% presented. Separate the keywords with commas.

% This command processes the author and affiliation and title information and builds
% the first part of the formatted document.
\maketitle
\vspace{-1mm}
{\fontsize{8pt}{8pt} \selectfont
\textbf{ACM Reference Format:}\\
Lu Cheng, Kai Shu, Siqi Wu, Yasin N. Silva, Deborah L. Hall, Huan Liu. 2020. Unsupervised Cyberbullying Detection via Time-Informed Gaussian Mixture Model. In \textit{Proceedings of the 29th ACM International Conference on Information and Knowledge Management (CIKM '20), October 19--23, 2020, Virtual Event, Ireland.} ACM, New York, NY, USA, 10 pages. \url{https://doi.org/10.1145/3340531.3411934}}

\section{Introduction}
\begin{figure}
\center
  \includegraphics[width=.8\columnwidth]{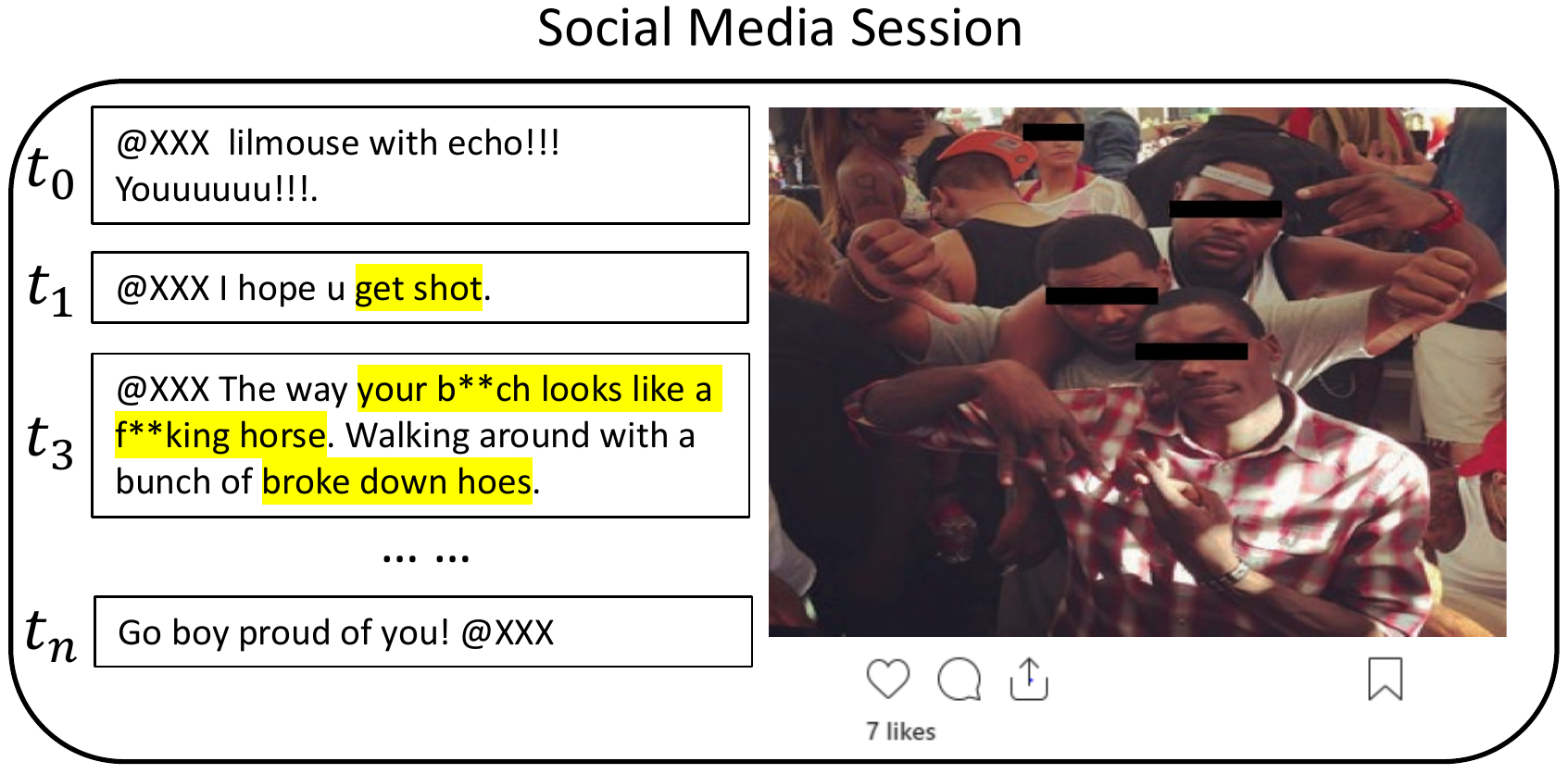}
  \caption{Illustration of a social media session on Instagram. Cyberbullying comments are repetitively posted by multiple users. Bullying words are highlighted. This work seeks to predict whether a given social media session is bullying.}
  \vspace{-2mm}
  \label{problem}
\end{figure}

Cyberbullying, defined as ``aggressively intentional acts carried out by a group or an individual using electronic forms of contact, \textit{repeatedly} or \textit{over time} against victims who cannot easily defend themselves''~\cite{smith2008cyberbullying}, has been rising at an alarming rate. Previous research has found that nearly 43\% of teens in the United States have been victims of cyberbullying~\cite{moessner2014cyberbullying}. In light of this, efforts aimed at automatically detecting cyberbullying -- which seeks to predict whether or not human interactions within a social media session constitute cyberbullying -- have a profound societal impact. However, detecting cyberbullying on social platforms is particularly challenging given that a social media session often consists of multi-modal information, for instance, an initial post, a sequence of comments, images/videos, and other social content such as the number of likes and shares. Fig. \ref{problem} illustrates an Instagram cyberbullying session where multiple bullying comments are posted.

Existing work on cyberbullying detection is mainly based on supervised methods, which often require a large annotated dataset for training. Although these approaches have shown promising results, they suffer from two major limitations: (1) Obtaining a large number of high-quality annotations for cyberbullying is time-consuming, labor-intensive, and error-prone because it requires circumspect examinations of multiple information sources such as images, videos, and numerous comments \cite{hosseinmardi2015analyzing}; (2) Current guidelines for labeling a session as cyberbullying may not be effective in the future due to the dynamic nature of language usage and social networks. Hence, we study alternative mechanisms for \textit{unsupervised} cyberbullying detection, which draws inferences from input social media data but without labeled responses.

Despite potential benefits, unsupervised cyberbullying detection also encounters several challenges: (1) Because cyberbullying typically consists of repetitive acts (as shown in Fig. \ref{problem}), the temporal dynamics of users' commenting behaviors adds nuanced understandings to the text-based methods that consider each comment as a distinct event over time. Such temporal characterization have been shown to be useful in distinguishing cyberbullying from non-bullying instances \cite{cheng2019hierarchical,guptatemporal,soni2018time}. Therefore, a key challenge is how to simultaneously model temporal dynamics and cyberbullying detection such that the two tasks mutually improve each other. (2) Social media sessions inherently present a \textit{hierarchical structure} where words form a comment and comments form a session. Previous studies \cite{yang2016hierarchical,cheng2019hierarchical} have revealed that modeling the hierarchical structure is useful for learning high-quality representations. Additionally, because meanings of words and comments are largely context-dependent, the sequential structure of words and comments need to be properly modeled for identifying relevant ones (e.g., the highlighted words in Fig. \ref{problem}); (3) A straightforward approach for unsupervised cyberbullying detection is to use the off-the-shelf clustering algorithms (e.g., $k$-means). The effectiveness of this approach largely relies on the quality of input data, however, social media data is notorious for its noise, sparsity, and high-dimensionality. Applying dimensionality reduction to the input data still presents the drawback of \textit{decoupled training}, i.e., representation learning and clustering are carried out separately.

To address these challenges, we propose a principled unsupervised learning framework -- \underline{U}nsupervised \underline{C}yberbullying \underline{D}etection via Time-Informed Gaussian Mixture Model (UCD). A central feature of UCD is that it incorporates the comment inter-arrival times of a social media session, which enables the classification of cyberbullying instances using the full commenting history. UCD consists of two main components: a \textit{representation learning} network, which learns the compact multi-modal representations of a session; and a \textit{multi-task learning} network, which predicts whether or not a session contains bullying behaviors while modeling the temporal dynamics of all comments. Specifically, the representation learning network models social media sessions using a Hierarchical Attention Network (HAN) \cite{yang2016hierarchical} for textual features and a Graph Auto-Encoder (GAE) \cite{kipf2016variational} for user and network features. The multi-task learning network then takes the multi-modal representations (e.g., text, user, and social network) as input to estimate the bullying likelihood using a time-informed Gaussian Mixture Model (GMM). The two UCD components are jointly optimized to mutually boost their learning effectiveness.

The main contributions of this paper are:
\begin{itemize}[leftmargin=*]
    \item We address the problem of unsupervised cyberbullying detection in social media platforms, which automatically identifies bullying instances without labeled data.
    \item We propose a principled framework for unsupervised cyberbullying detection, which includes two components that jointly learn low-dimensional representations and predict bullying instances.
    \item We conduct experiments on two real-world social datasets from Instagram and Vine. Our results show that UCD not only outperforms the state-of-the-art unsupervised models, but also achieves competitive performance against supervised models\footnote{Code available at \url{https://github.com/GitHubLuCheng/UCD}}.
\end{itemize}
% The remainder of this paper is organized as follows: we review the related work in Sec. 2, and describe the details of the proposed UCD framework in Sec. 3. In Sec. 4, we present the quantitative and qualitative evaluation of the proposed approach, followed by a discussion of the practical impact of UCD in Sec. 5. We summarize our findings and discuss directions for future work in Sec. 6.
\section{Related Work}
We review related work on automatic cyberbullying detection models and clustering algorithms based on deep neural networks.    
\subsection{Cyberbullying Detection}
To date, cyberbullying has received a significant amount of attention within psychology and social science fields. It has only more recently become a focus of computer science research, where much of the work has been aimed at developing models that automatically identify bullying behaviors. For instance, existing work on automatic cyberbullying detection has used manually labeled data to mine patterns from text \cite{dinakar2011modeling,nand2016bullying,xu2012learning,dani2017sentiment,cheng2019hierarchical,romsaiyud2017automated}, social network \cite{chatzakou2017mean,huang2014cyber,liu2018forecasting}, and other media sources such as images and videos \cite{hosseinmardi2015analyzing,hosseinmardi2016prediction,rafiq2015careful,rafiq2016analysis,cheng2019xbully}. Xu et al. \cite{xu2012learning} explored several natural language processing (NLP) techniques to identify bullying traces and further defined the structure of a bullying episode and the associated roles (e.g., victims and bullies) on Twitter. Dinakar et al. \cite{dinakar2011modeling} concatenated TF-IDF scores, POS tags of frequent bigrams, and profane words as content features to detect cyberbullying on a manually-labeled corpus of YouTube comments. Dani et al. \cite{dani2017sentiment} sought to incorporate sentiment into the content features by capturing the sentiment consistency of bullying and non-bullying posts. Most recently, Ziems et al. \cite{ziems2020aggressive} characterized cyberbullying using five explicit factors to represent its social and linguistic aspects.

Although many researchers define cyberbullying as a harmful behavior that is repeated over time, relatively little work has examined the temporal aspects of cyberbullying. Among the few studies that have, Soni and Singh \cite{soni2018time} modeled the commenting behaviors as Poisson point processes and identified several temporal features that help distinguish bullying sessions. Cheng et al. \cite{cheng2019hierarchical} employed a hierarchical attention network to capture the sequence-aware structure of words and comments in a social media session and integrated time interval prediction into the detection model. From a causality perspective, Cheng et al. \cite{cheng2019robust} sought to discover the potential confounders among bullying texts so that the resulting classifiers can be transferred between different domains.

Crucially, most existing work on cyberbullying detection has focused on supervised learning models that require large-scale labeled datasets. To reduce this dependency on human-coded data, Raisi and Huang \cite{raisi2017cyberbullying} proposed a weakly-supervised model that starts with a small seed vocabulary of bullying indicators. They then extracts bullying roles and additional bullying indicators based on an unlabeled corpus of social media interactions. Another work \cite{raisi2017co} studied cyberbullying detection with an ensemble of two learners that co-train one another; one learner examines the language content in the messages while the other considers the social structure. To our knowledge, the only unsupervised cyberbullying detection model GHSOM \cite{di2016unsupervised} inputs several NLP and social features into the Growing Hierarchical SOM using the SOMToolbox framework.

\subsection{Deep Clustering} Clustering methods based on deep neural networks have shown promising results in real-world applications (e.g., anomaly detection \cite{zong2018deep}) due to their high representational power. Standard clustering-friendly representations are learned with a two-phase training procedure. In the first phase, the auto-encoder is trained with the mean squared error reconstruction loss. In the second phase, the auto-encoder is further fine-tuned with a combined loss function consisting of the reconstruction loss and a clustering-specific loss.
For example, Song et al. \cite{song2013auto} applied an auto-encoder in the clustering tasks and introduced a new objective function that includes the reconstruction error and the distance between data and their corresponding cluster centers in the latent space. Similarly, the Deep Embedded Clustering model in \cite{xie2016unsupervised} projected the data from an original space to a lower-dimensional feature space and then jointly optimized a clustering objective using stochastic gradient descent via backpropagation. Relevant to the present work, multiple studies have employed unsupervised anomaly detection \cite{chandola2009anomaly}. For instance, Zong et.al \cite{zong2018deep} jointly optimized the parameters of a deep auto-encoder and a mixture model with the two components mutually improving each other's performance.

In contrast to most existing cyberbullying detection models, UCD focuses on the unsupervised approach where labeled data is not available during training. To achieve good performance, we exploit multi-modal data and relevant information such as the temporal patterns of comments and the hierarchical structure of social media sessions. Our evaluation results show that the integration of this additional information can significantly improve the effectiveness of unsupervised cyberbullying detection.
\begin{figure*}
\center
  \includegraphics[width=.9\textwidth]{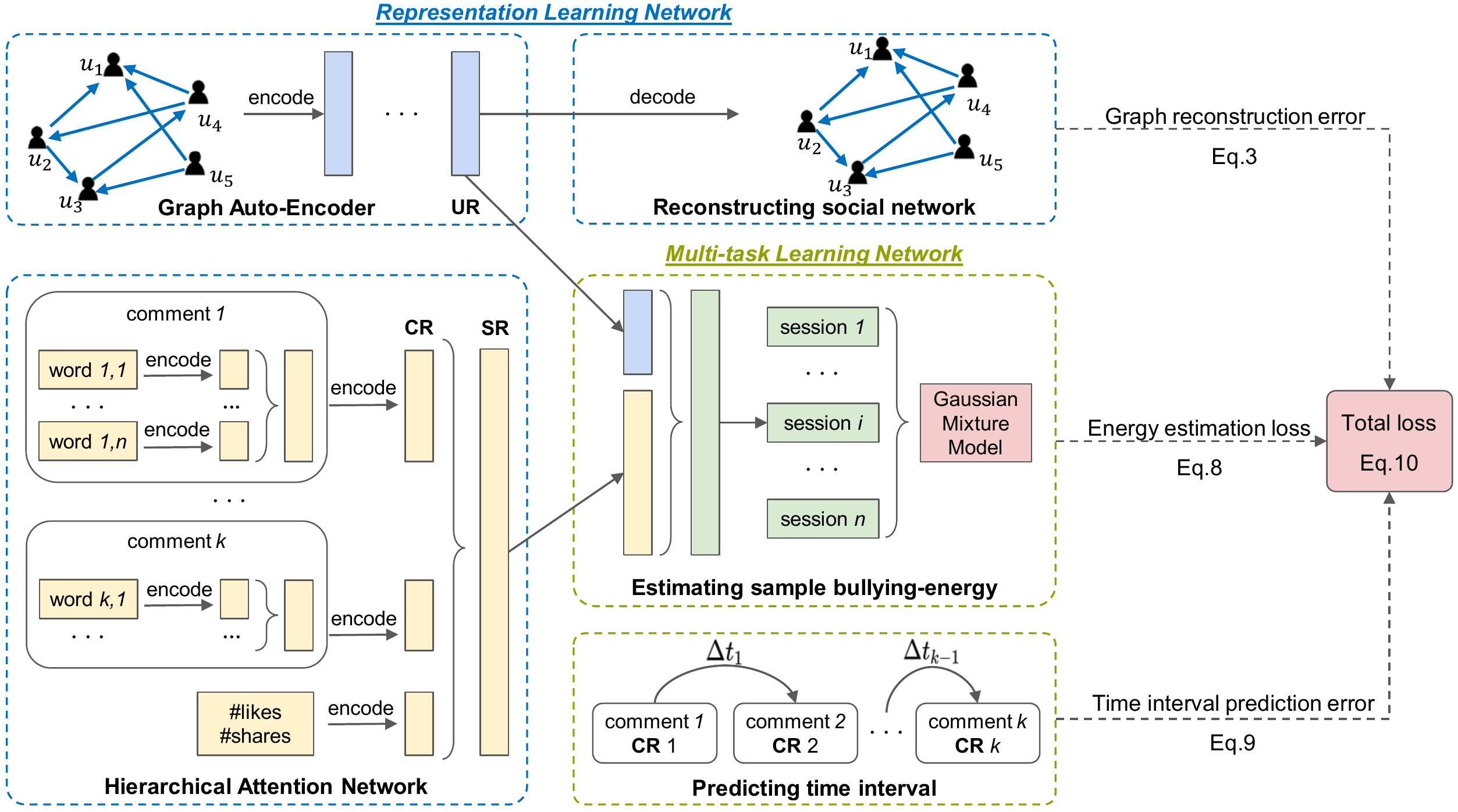}
  \caption{Overview of the proposed framework. UCD consists of two components: (1) The \textit{representation learning network} (the blue dashed rectangles) constructs multi-modal representations of social media sessions (the green solid rectangles in the middle); (2) The \textit{multi-task learning network} (the green dashed rectangles) that simultaneously estimates the energy/likelihood of input samples and predicts time intervals between comments. Observe that the representation learning network combines user (session owner) representation (\textbf{UR}) in the Graph Auto-Encoder (top part) and social representation (\textbf{SR}) in the Hierarchical Attention Network (HAN, bottom left) to form the session representation. The constructed session representation is the input of the sample bullying energy estimation task. Meanwhile, the comment representations (\textbf{CR}) in HAN are fed into the time interval prediction task (bottom part). The overall loss comes from three sources: graph reconstruction error, energy estimation loss, and time interval prediction error. Best viewed in colors.}
  \label{framework}
\end{figure*}
\section{UCD: The Proposed Framework}
The framework overview in Fig. \ref{framework} shows that our model consists of two major components: (1) a representation learning network that leverages HAN and GAE to obtain multi-modal representations, and (2) a multi-task learning network that jointly optimizes a GMM-based energy estimation task to detect cyberbullying instances and a temporal prediction task to further refine the session representations with the comment inter-arrival times.

\subsection{Representation Learning Network}
Social media sessions usually consist of multi-modal information, such as text (e.g., comments) and social content (e.g., friendship networks, number of likes and shares). The representation learning network aims to transform these sparse and high-dimensional features into a low-dimensional session representation.

\noindent{\textbf{HAN for Text.}} The majority of prior literature on cyberbullying detection considered the comments in a social media session as independent events and directly extracted textual features from a chunk of combined comments. Notwithstanding its simplicity, this method largely overlooks the hierarchical structure of a social media session and the long-term dependencies among the sequentially posted comments. Previous studies showed that i) modeling document structure can significantly improve the quality of document representations \cite{yang2016hierarchical}; and ii) capturing long-term dependencies is particularly useful for sequential data modeling \cite{dietterich2002machine}. In addition, different words and comments in a post are not equally relevant for cyberbullying detection, i.e., some words/comments are more important than others. For example, \textit{``You're a f**king loser!''} and \textit{``Yeah, I'm a loser.''} both include the word \textit{loser}, the former is, however, more likely to represent an instance of bullying. Therefore, we also integrate attention mechanisms to distinguish important words and comments. Following \cite{cheng2019hierarchical}, we employ a hierarchical attention network to generate the textual representation for a social media session. The HAN approach is a particularly good fit in cyberbullying detection as it models the two main levels of social media sessions (sequences of words and comments) and at each level, the model captures the long-term dependencies and integrates mechanisms to differentiate the importance of specific words and comments based on their context.

The hierarchical structure of the textual content can be described as follows: a social media session consists of a sequence of comments and each comment includes a sequence of words. Given a session with $C$ comments where each comment $i$ has $L_i$ words $\{w_{it}|t=1,2,...,L_i\}$, we use the bi-directional Gated Recurrent Units (GRUs) \cite{bahdanau2014neural} to model both the word sequence in a comment and the comment sequence in a session:
\begin{align}
\begin{split}
    \overrightarrow{s}_{it}=\overrightarrow{GRU}(W_ew_{it}),\quad \forall t\in [1,L_i], i \in [1,C]\\
    \overleftarrow{s}_{it}=\overleftarrow{GRU}(W_ew_{it}),\quad \forall t\in [L_i,1], i \in [1,C]
\end{split}
\end{align}
where each word $w_{it}$ is first mapped to a latent space with parameter $W_e$. The resulting annotation for word $w_{it}$ is a concatenation of the forward and backward hidden states, $s_{it}=[\overrightarrow{s}_{it},\overleftarrow{s}_{it}]$. To differentiate the word importance, we adopt the attention mechanism \cite{bahdanau2014neural,yang2016hierarchical} to automatically detect words that are more relevant and then aggregate the representation of weighted words to form a comment vector $c_i$:
\begin{equation}
    \alpha_{it}=\frac{\exp(h_{it}^Tu_w)}{\sum_t\exp(h_{it}^Tu_w)}; \quad c_i=\sum_t\alpha_{it}s_{it},
\end{equation}
where $h_{it}$ is the output of a fully connected layer of $s_{it}$ and $u_w$ denotes a word-level context vector \cite{yang2016hierarchical}. $\alpha_{it}$ denotes a normalized weight describing the importance of word $w_{it}$. Similarly, the final textual representation $v$ of a social media session can be computed using the encoded comment vectors (i.e., replacing $w_{it}$ of Eq.1 with $c_{i}$). Further, we include a dense layer to project the social content, i.e., number of likes and shares, into a latent space. We later concatenate the resulting vector $p$ with $v$ to form the multi-modal representation of a social media session $o=[v,p]$.

\noindent{\textbf{GAE for Attributed Social Networks.}} Self-selection bias (grouping with similar others) and peer influence are closely connected with bullying behaviors in offline environments~\cite{espelage2003examination,witvliet2010peer,salmivalli1997peer,chengpi}. Research in human communication \cite{festl2013social} reveals a similar observation that online social network positioning is a comparably strong predictor for cyberbullying detection. Hence, it is important to consider the social network structure and peer influence from similar users for improving the performance of cyberbullying detection.

The representation learning network learns user representation by exploiting information from social networks where nodes denote social media users with corresponding profile information being the node attributes, and edges denote the follower/followee relationships. Here, we employ GAE to embed users' attributes as low-dimensional vectors such that users with structural proximity in the social network are close. As one of the most powerful node embedding approaches, GAE has been applied to several challenging learning tasks such as link prediction \cite{grover2018graphite,kipf2016variational} and node clustering \cite{salha2019degeneracy}. GAE can effectively incorporate node features and learn more interpretable user representations \cite{kipf2016variational}. The key of GAE is the encoding-decoding scheme, i.e., GAE encodes nodes into low-dimensional vectors which are then decoded to reconstruct the original network structure. Suppose we are given a social network $\mathcal{G}=(\mathcal{V},\mathcal{E})$ with $U=|\mathcal{V}|$ users. The adjacency matrix of this graph is $A \in \mathbb{R}^{U\times U}$. The User-Feature matrix is $X \in \mathbb{R}^{U\times D}$ with $D$ being the feature dimension. GAE then uses a graph convolutional network (GCN) \cite{kipf2016semi} encoder and an inner product decoder to learn a latent matrix $Z$ by minimizing the following reconstruction error:
\begin{equation}
\begin{aligned}
     g&=\frac{1}{2}\|A-\hat{A}\|_2^2, \\ 
    \text{ with }\hat{A}=&\sigma(ZZ^T), Z= \text{GCN}(X,A)
\end{aligned}
\end{equation}
where $\sigma(\cdot)$ is the logistic sigmoid function. 

The final representation of a session is the concatenation of user (owner) representation and the representation output from HAN, i.e., $ss=[z,o]$, where $z$ is a row vector of $Z$. This multi-modal representation is then fed into the multi-task learning network.
\subsection{Multi-Task Learning Network}
Given the multi-modal representation of input sessions, the multi-task learning network simultaneously (1) estimates the sample bullying-energy/likelihood; and (2) models the inter-arrival times of a sequence of comments in a social media session. These two tasks can mutually enhance each other's performance in the training stage. To this end, the multi-task learning network enables the proposed framework to jointly learn session representations and discover cyberbullying instances.

\noindent{\textbf{Bullying-energy estimation.}} The first task of the multi-task learning network is to estimate the sample energy (likelihood) and classify samples with high energy (low likelihood) as bullying instances. A primary benefit of energy-based models is the flexibility to specify the energy expression \cite{zhai2016deep}. Here, we construct a GMM-based density estimator to infer the underlying probability density function. GMM, a widely used unsupervised learning method, seeks to fit a multi-modal distribution with multiple unimodal Gaussian distributions which are the most commonly used distributions for modeling real-world unimodal data. Previous work \cite{zong2018deep,zhai2016deep} has shown that GMM is more effective than simple models for data with complex structures. Given the complexity and multi-modal nature of social media data, we leverage GMM to perform density estimation tasks over multi-modal representations.

Let the number of mixture components be $K$ and the latent representation of a social media session be $ss$, we first generate the mixture membership predictions for $ss$. We then estimate the parameters of GMM using the predicted membership to obtain the energy estimation of $ss$. Specifically, we first feed $ss$ into a multi-layer network (MLN) \cite{svozil1997introduction} parameterized by $\theta_m$. The output is denoted as $p_{MLN}$:
\begin{equation}
    p_{MLN}=\text{MLN}(ss;\theta_m)
\end{equation}
The probability of $ss$ belonging to each component can be estimated as follows:
\begin{equation}
    \hat{m}=\text{softmax}(p_{MLN})
\end{equation}
where $\hat{m}$ is a $K$-dimensional vector. Given a batch of $N$ social media session representations $\{ss_1,ss_2,...,ss_N\}$, together with the corresponding predicted memberships, we can further estimate the parameters in GMM as follows:
\begin{equation}
    \hat{\phi}_k=\sum_{i=1}^N \frac{\hat{m}_{ik}}{N}; \quad
    \hat{\mu}_{k}=\frac{\sum_{i=1}^N\hat{m}_{ik}ss_i}{\sum_{i=1}^N\hat{m}_{ik}}
\end{equation}
\begin{equation}
\hat{\Sigma}_k=\frac{\sum_{i=1}^N\hat{m}_{ik}(ss_i-\hat{\mu}_k)(ss_i-\hat{\mu}_k)^T}{\sum_{i=1}^N\hat{m}_{ik}}
\end{equation}
where $\hat{\phi}_k$, $\hat{\mu}_k$ and $\hat{\Sigma}_k$ denote the mixture probability, mean, and covariance of component $k \in \{1,2,...,K\}$ in GMM, respectively. $\hat{m}_{ik}$ denotes the probability of $ss_i$ in the $k$-th component of GMM. To build the probability density function, we leverage the energy-based model \cite{lecun2006tutorial} which relies on a specific parameterization of the energy (negative log likelihood).
The energy level of a session is defined as:
\begin{equation}
    E(ss_i;\theta_m)=-\log \Big(\sum_{k=1}^K\hat{\phi}_k\frac{\exp\big(-\frac{1}{2}(ss_i-\hat{\mu}_k)^T\hat{\Sigma}_k^{-1}(ss_i-\hat{\mu}_k)\big)}{\sqrt{|2\pi\hat{\Sigma}_k|}}\Big)
\end{equation}
where $|\cdot|$ is the determinant of a matrix. The model then classifies a session as cyberbullying if its energy is above a predefined threshold $\tau \in$ (0,1) in the testing phase. In practice, $\tau$ is typically set to a comparatively large value, i.e., a cyberbullying session is in general associated with high energy (hence low likelihood). This is because bullying samples are less frequently observed in real-world datasets, as suggested by the statistics in Table \ref{data} as well as in previous literature \cite{chengpi,dinakar2012common}. 

\noindent{\textbf{Temporal dynamics fitting.}} Cyberbullying is commonly defined as a \textit{repeated act} of aggression that develops over time \cite{soni2018time,cheng2019hierarchical,dinakar2012common}. However, most of the existing computational models consider each comment in a social media session as an isolated event. Therefore, they largely overlook the temporal dynamics of users' commenting behavior. Here, we seek to predict the inter-arrival times between comments for obtaining additional feedback from the temporal dynamics. This feature enables the model to exploit the commonalities and differences across bullying-energy estimation and temporal-dynamics prediction for improving the final cyberbullying detection performance.

We first obtain the output $e_{in}$ of the comment encoder for comment $i$ in session $n$ from the HAN module and then conduct a time interval prediction task as follows.
\begin{equation}
    \ell=\sum_{i=1}^C\frac{1}{2}\|f(e_{in}; \theta_\ell) - \Delta t_i \|^2,
\end{equation}
where $f$ represents a regression model, $\theta_\ell$ denotes the associated parameters, and $\Delta t_i=t_i-t_{i-1}$ is the time interval between comment $i-1$ and $i$. We set $t_0$ to be 0. Let $d$ denote the dimensions of the latent representation of social media sessions, $\theta_h$ the parameters of HAN and $\theta_g$ the parameters of GAE, the final objective function of UCD can be constructed as:
\begin{align}
\begin{split}
    J=\sum_{n=1}^N\sum_{i=1}^C\frac{1}{2}\|f(e_{in}; \theta_\ell) - \Delta t_i \|^2+\frac{\lambda_1}{N}\sum_{i=1}^NE(ss_i;\theta_m)\\
    +\frac{\lambda_2}{2}\|A-\hat{A}\|^2_2+\lambda_3P(\hat{\Sigma}); \text{ with } P(\hat{\Sigma})=\sum_{k=1}^K\sum_{j=1}^d\frac{1}{\hat{\Sigma}_{kjj}}
\end{split}
\label{obj}
\end{align}
$P(\hat{\Sigma})$ accounts for the singularity issue in GMM, $\lambda_1, \lambda_2,$ and $\lambda_3$ are the hyperparameters that control the balance among time interval prediction error, energy estimation loss, graph reconstruction error and regularization for GMM. Specifically, the objective function consists of four components (ordered as presented in Eq. \ref{obj}):
\begin{itemize}[leftmargin=*]
    \item The first component is the loss function that describes the prediction error of time interval prediction. 
    \item The second component $E(ss_i;\theta_m)$ models the likelihood (sample energy) that session $i$ is observed. Here, minimizing the energy level of an input session will maximize the likelihood of observing the session. 
    \item The third component is the reconstruction error of GAE in the representation learning network. A lower error indicates that the learned user representations better preserve the structure of the original attributed social network. 
    \item Due to the singularity issue in GMM, we penalize small values on the diagonal entries of the covariance matrices $\hat{\Sigma}$.
\end{itemize}

The proposed model jointly optimizes the representation learning network and the multi-task learning network to learn effective representations for cyberbullying detection. We train the model by minimizing Eq. \ref{obj} using the Adam optimization algorithm \cite{kingma2014adam}, where the error backpropagates through the representation learning network, the bullying-energy estimation task, and the time-interval prediction task.

\section{Evaluation}
In this section, we present both quantitative and qualitative analyses to evaluate the proposed UCD framework. 
Specifically, we answer the following research questions: \\
(1) \textit{Effectiveness}: \textbf{a.} How effective is UCD compared to existing unsupervised learning approaches and supervised classification models? \textbf{b.} How does each module, i.e., HAN, GAE, and temporal modeling, affects the cyberbullying detection performance of UCD? \\
(2) \textit{Robustness}: How robust is UCD when varying model parameters? 
% ==============================

\subsection{Datasets}
Our experiments use two public datasets crawled from Instagram\footnote{https://www.instagram.com/} and Vine\footnote{https://vine.co/} (now in archive status). The datasets were introduced and released in \cite{hosseinmardi2015analyzing} and \cite{rafiq2015careful}, respectively. The basic statistics of these datasets are presented in Fig. \ref{data}. \\
\textbf{Instagram}: Instagram is a popular social media platform. It is also the platform on which the highest prevalence of cyberbullying has been reported \cite{Ditch2013}. 
Using a snowball sampling method, the authors in \cite{hosseinmardi2015analyzing} identified 41K Instagram users, 61\% of whom had public profiles. For each public user, the collected data includes the media objects the user had posted, the comments of session, the list of user followers/followees, and the list of users who have commented/liked the media objects. Data labeling (whether the session constituted cyberbullying or not) was conducted on CrowdFlower\footnote{http://www.figure-eight.com/} -- a crowdsourcing website -- using a procedure whereby each session was labeled by five different contributors. A session is labeled as cyberbullying if three or more contributors had labeled this session as cyberbullying. Overall, the \textit{Instagram} dataset includes 2,218 labeled social media sessions. \\
\textbf{Vine}: The Vine dataset \cite{rafiq2015careful} is used for analyzing cyberbullying in the context of a video-based online platform. 
It was crawled using a snowball sampling method in which a random user $u$ is first selected as a seed and then the crawling continues with the users that $u$ follows. Each session includes videos, captions, and associated comments (note that social network information was not available for this dataset). All sessions in the dataset have at least 15 comments. Similar to the labeling process used for the \textit{Instagram} data, a total of 970 \textit{Vine} sessions were labeled (as cyberbullying vs. non-bullying) using CrowdFlower.

\begin{table}
\begin{center}
\caption{Basic statistics for \textit{Instagram} and \textit{Vine} datasets.}
\begin{tabular}{ c||c|c|c|c } \hline
Datasets  & \#Sessions &\#Bully &\#Non-bully &\#Comments \\ \hline
\textit{Instagram} & 2,218 & 678 & 1,540 & 155,260 \\ \hline
\textit{Vine}&  970 & 304 & 666& 78,250 \\ \hline
\end{tabular}
\label{data}
\end{center}
\end{table}
We use the following information gathered from a media session: 
\begin{itemize}[leftmargin=*]
    \item Attributed social network: A social network where each node represents a user and has attributes such as the number of total followers and followees. The edges denote the following and followed-by relationships.
    \item Text: The bag-of-words representation of the captions and comments. Each column indicates a term from the corpus and the entry is the corresponding frequency count.
    \item Time: The posting timestamps of a media object and its associated comments. We extract the time difference between any two consecutive comments.
    \item Social content: The number of likes and shares of a post receives.
\end{itemize} 
\subsection{Experimental setup}
To answer the first research question, we compare UCD with multiple unsupervised learning models:
\begin{itemize}[leftmargin=*]
    \item \textbf{$k$-means.} $k$-means is one of the most common clustering algorithms. It iteratively assigns each data point to one of $k$ groups with the smallest distance.
    \item \textbf{HAE} \cite{li2015hierarchical}. HAE is an LSTM model that hierarchically builds embeddings for social media sessions from comments and words. We also used $k$-means to cluster the learned representations.
    \item \textbf{DCN} \cite{yang2017towards}. DCN is a deep learning-based clustering algorithm that regulates auto-encoder performance by using $k$-means.
    \item \textbf{DAGMM} \cite{zong2018deep}. DAGMM jointly optimizes a deep auto-encoder that learns low-dimensional representations and a GMM that estimates the density function of the latent representations.
    \item \textbf{XBully} \cite{cheng2019xbully}. XBully learns multi-modal representations of social media sessions and then feeds them into a subsequent classification model. We replaced the classification model with $k$-means.
    \item \textbf{GHSOM} \cite{di2016unsupervised}. To our knowledge, Growing Hierarchical Self-Organizing Map (GHSOM) is the only existing model for unsupervised cyberbullying detection. It extracts sentiment, syntactic, and semantic features from text and social network data. The features are then fed into the GHSOM tool\footnote{http://www.ifs.tuwien.ac.at/~andi/ghsom/} for clustering.
\end{itemize}

To provide a comprehensive analysis of UCD, we also include the following supervised methods:
\begin{itemize}[leftmargin=*]
    \item \textbf{Na\"ive Bayes (NB)}. NB is a probabilistic classifier based on Bayes' theorem with strong independence assumptions between the features. It is one of the most popular (baseline) methods for text classification.
    \item \textbf{Random Forest (RF)}. RF consists of several individual decision trees that operate as an ensemble. Each individual tree generates a class prediction and the class with the most votes becomes the model's prediction.
    \item \textbf{Logistic Regression (LR)}. LR is a statistical model that uses a logistic function to model a binary dependent variable. It is a common baseline algorithm for binary classification.
\end{itemize}
For baselines using $k$-means, we set the number of clusters to 2, and label the cluster with fewer elements as \textit{bullying} and the other one as \textit{non-bullying}. This assumption is supported by the statistics in Table \ref{data} and also generally evident in other real-world cyberbullying datasets~\cite{ziems2020aggressive}.  Note that our proposed method (UCD) does not require this assumption as it optimizes Eq. \ref{obj} for clustering bullying and non-bullying instances. We implemented the following variants of UCD to examine the impact of each UCD component.
\begin{itemize}[leftmargin=*]
    \item \textbf{UCDXtext.} UCD without HAN. We do not report this variant for \textit{Vine} given that its social network information is not available.
        \item \textbf{UCDXtime.} UCD without time interval prediction.
    \item \textbf{UCDXgraph.} UCD without GAE.
\end{itemize}
Following previous literature \cite{soni2018time,chengpi}, we use four common evaluation metrics -- \textit{Precision}, \textit{Recall}, \textit{F1}, and \textit{AUROC (Area Under the Receiver Operating Characteristic Curve)}. Note that we are more interested in detecting cyberbullying instances, therefore, we report Precision, Recall and F1 corresponding to the bullying (positive) class. While the overall performance can be effectively measured by F1 and AUROC scores, multiple application scenarios of cyberbullying detection could particularly benefit from the identification of as many positive cases as possible, i.e., high Recall. \\
\noindent\textbf{Parameter Setting.}
Based on Eq. \ref{obj}, the UCD framework has five hyperparameters: (1) $\lambda_1$, for balancing the sample bullying-energy loss; (2) $\lambda_2$, for controlling the weight of the reconstruction error of GAE; (3) $\lambda_3$, for controlling the weight of diagonal entries in the covariance matrices; (4) $K$,\footnote{This is different from the $k$ in $k$-means, which sets the number of clusters (bullying and non-bullying). $K$ in GMM denotes the number of memberships and relates to computing the sample energy. We use the energy threshold to detect bullying instances.} the number of mixtures in the GMM; and (5) $\tau \in(0,1)$, a pre-defined energy threshold. We set the parameters based on sensitivity analysis, which is detailed in Section 4.5. Specifically, we set $\lambda_1=1e-4$, $\lambda_3=1e-9$ and $K=5$ for both datasets. The energy threshold $\tau$ is set to 65\% for \textit{Instagram} and 70\% for \textit{Vine}. Therefore, Instagram and Vine test sessions with the highest 35\% and 30\% energy values will be classified as bullying cases and the rest as non-bullying cases, respectively. For \textit{Instagram}, we additionally set $\lambda_2=0.01$. For the baseline methods, we conducted similar sensitivity analysis on the key parameters reported in their original papers. For both datasets, we use 80\% of the data for training and the rest for testing. Each experiment is run 10 times, mean and standard deviations are reported.
\begin{table}
\small
\setlength{\tabcolsep}{4pt} % Default value: 6pt
\centering
\caption{Performance evaluation with \textit{Instagram} data.}
\begin{tabular}{|c||c|c|c|c|}
\hline
\multicolumn{5}{|c|}{\textit{Unsupervised Learning Models}}\\\hline
Metrics & Precision& Recall & F1& AUROC \\ \hline 
$k$-means  &0.79$\pm$0.02 &0.29$\pm$0.04 &0.43$\pm$0.05&0.63$\pm$0.02 \\ \hline
XBully &0.32$\pm$0.02&0.47$\pm$0.03& 0.38$\pm$0.02&0.51$\pm$0.02 \\\hline
HAE &0.53$\pm$0.02&0.27$\pm$0.03& 0.35$\pm$0.03&0.53$\pm$0.01 \\ \hline
DCN &\textbf{0.87$\pm $0.02}& 0.23$\pm$0.02 & 0.36$\pm$0.02&0.61$\pm$0.01\\ \hline
DAGMM &0.56$\pm$0.18&0.56$\pm$0.18& 0.56$\pm$0.18&0.56$\pm$0.03  \\ \hline
GHSOM   & 0.35$\pm$0.12 &0.38$\pm$0.06&0.36$\pm$0.08&0.54$\pm$0.11\\ \hline
UCDXtext &0.33$\pm$0.01&0.34$\pm$0.01& 0.33$\pm$0.01&0.53$\pm$0.02 \\ \hline
UCDXtime &0.47$\pm$0.02&0.48$\pm$0.01& 0.48$\pm$0.01&0.63$\pm$0.01 \\ \hline
UCDXgraph &0.56$\pm$0.02& 0.57$\pm$0.01 & 0.57$\pm$0.02 & 0.69$\pm$0.01  \\ \hline
UCD   &0.59$\pm$0.02 &\textbf{ 0.66$\pm$0.02} & \textbf{0.63$\pm$0.02}&\textbf{0.73$\pm$0.01} \\ \hline
\multicolumn{5}{|c|}{\textit{Supervised Learning Models}}\\\hline
Metrics & Precision& Recall & F1& AUROC \\ \hline 
NB &0.40$\pm$0.03&\textbf{0.69$\pm$0.03}&0.51$\pm0.03$&0.62$\pm$0.02\\ \hline
RF& 0.78$\pm$0.03&0.53$\pm$0.03&0.63$\pm$0.03&0.73$\pm$0.01\\\hline
LR&\textbf{0.79$\pm$0.03}&0.55$\pm$0.03&\textbf{0.64$\pm$0.03}&\textbf{0.74$\pm$0.03}\\\hline
\end{tabular}
\label{table:Ins}
\end{table}
% ==============================
%=================================
\begin{table}
\small
\setlength{\tabcolsep}{4pt} % Default value: 6pt
\centering
\caption{Performance evaluation with \textit{Vine} data.}
\begin{tabular}{|c||c|c|c|c|}
\hline
\multicolumn{5}{|c|}{\textit{Unsupervised Learning Models}}\\\hline
Metrics & Precision& Recall & F1& AUROC \\ \hline 
$k$-means&0.03$\pm$0.08 &0.00$\pm$0.00 &0.00$\pm$0.01&0.50$\pm$0.00 \\\hline
XBully&\textbf{0.48$\pm$0.08}&0.27$\pm$0.03& 0.34$\pm$0.04 & 0.57$\pm$0.02 \\\hline
HAE&0.18$\pm$0.04 & 0.34$\pm$0.08 & 0.23$\pm$0.04 & 0.57$\pm$0.03 \\\hline
DCN &0.29$\pm $0.20& 0.32$\pm$0.39 & 0.22$\pm$0.19&0.50$\pm$0.03\\ \hline
DAGMM & 0.36$\pm$0.09 & 0.31$\pm$0.08& 0.33$\pm$0.08&0.54$\pm$0.00  \\ \hline
GHSOM   & 0.32$\pm$0.09 & 0.38$\pm$0.10 &0.34$\pm$0.08&0.50$\pm$0.07\\ \hline
UCDXtime&0.33$\pm$0.02&0.39$\pm$0.03& 0.36$\pm$0.02&0.56$\pm$0.01 \\ \hline
UCDXgraph &0.43$\pm$0.02&\textbf{0.40$\pm$0.03}& \textbf{0.41$\pm$0.02}&\textbf{0.58$\pm$0.01}\\\hline
\multicolumn{5}{|c|}{\textit{Supervised Learning Models}}\\\hline
Metrics & Precision& Recall & F1& AUROC \\ \hline 
NB&0.49$\pm$0.05&\textbf{0.72$\pm$0.05}&0.58$\pm$0.04&0.70$\pm$0.04\\\hline
RF&\textbf{0.67$\pm$0.05}&0.42$\pm$0.05&0.51$\pm$0.04&0.66$\pm$0.02\\\hline
LR&0.62$\pm$ 0.05&0.57$\pm$0.05&\textbf{0.59$\pm$0.04}&\textbf{0.71$\pm$0.03}\\\hline
\end{tabular}
\label{table:Vine}
\end{table}
\subsection{Quantitative Results}
For the \textit{Instagram} dataset, we compare UCD and its variants with all baselines. Due to the lack of social network information in the \textit{Vine} dataset, UCD and UCDXtext cannot be evaluated with \textit{Vine}. The best results for unsupervised and supervised models are highlighted in Table \ref{table:Ins} and \ref{table:Vine} with bold text. The results we present for RF are different from those reported in \cite{rafiq2016analysis}. We believe this is the case because the original work: 1) considered additional features such as the percentage of negative comments, emotions exhibited in videos, and latent semantic features (10 topics based on the comments using LDA), and 2) performed oversampling (SMOTE \cite{chawla2002smote}) to balance the \textit{Vine} dataset. We use the original \textit{Vine} dataset to better reflect real-world scenarios.

We observe that (1) UCD achieves the best performance in Recall, F1, AUROC, and competitive Precision compared to the unsupervised baselines for both datasets. For the \textit{Instagram} dataset, UCD shows 15.9\%, 19.7\%, and 35.2\% of improvement on AUROC compared to the results using raw features (i.e., $k$-means), representation learning (i.e., DCN), and the unsupervised cyberbullying detection model GHSOM, respectively. AUROC considers all possible thresholds for classification and is a more appropriate metric when datasets are imbalanced; (2) Imbalanced datasets affect the trade-off between Recall and Precision. While achieving superior Precision, baseline models DCN and $k$-means show poor Recall. We infer that these models fail to identify most of the cyberbullying instances, which is undesired in many cyberbullying applications; and (3) UCD achieves competitive Recall, F1 and AUC scores compared to supervised methods using the \textit{Instagram} dataset. For instance, LR improves F1 by 1.6\% over UCD whereas NB is outperformed by UCD regarding these three metrics. The Precision of UCD is comparatively low implying that its energy threshold favors identifying cyberbullying instances, therefore, UCD miss-classifies more non-bullying instances than baseline methods. In the \textit{Vine} dataset, the supervised methods show larger advantages over UCDXgraph, reflecting the importance of integrating social network information and using larger datasets in order to maximize the performance of UCD. Of particular interest is that UCD also achieves more balanced Precision and Recall values compared to supervised models.
% For example, RF presents much better Precision than Recall while NB shows superior Recall over Precision.

We make the following observations when comparing UCD with its own variants: (1) UCD achieves better performance in all metrics, especially against UCDXtext and UCDXtime, leading us to conclude that each submodule (HAN, GAE, and temporal analysis) has a positive influence on UCD's performance; (2) The performance of UCDXtext drops significantly compared to other variants, highlighting the importance of textual features in cyberbullying detection; (3) UCDXgraph outperforms UCDXtime, indicating that temporal analysis can provide more relevant information for cyberbullying detection than social network properties and thus highlighting the importance of modeling temporal patterns; and (4) the proposed framework performs better on \textit{Instagram} data than on \textit{Vine} data. This is in part due to the smaller sample size and lack of social network information in the \textit{Vine} dataset.

In summary, UCD outperforms unsupervised baselines in terms of identifying cyberbullying instances and the overall performance. Compared to supervised models, it shows competitive performance when the sample size is comparatively large and the social network information is available. None of the evaluated methods achieves high performance in detecting both bullying and non-bullying instances. Future work is encouraged to investigate such methods.
%Original text: UCD and baselines cannot achieve balanced performance in detecting both bullying and non-bullying instances. Future work is encouraged to investigate such methods.
\begin{figure*}
\captionsetup{justification=centering}
\centering
    \begin{subfigure}{.25\textwidth}
    \centering
      \includegraphics[width=.9\linewidth]{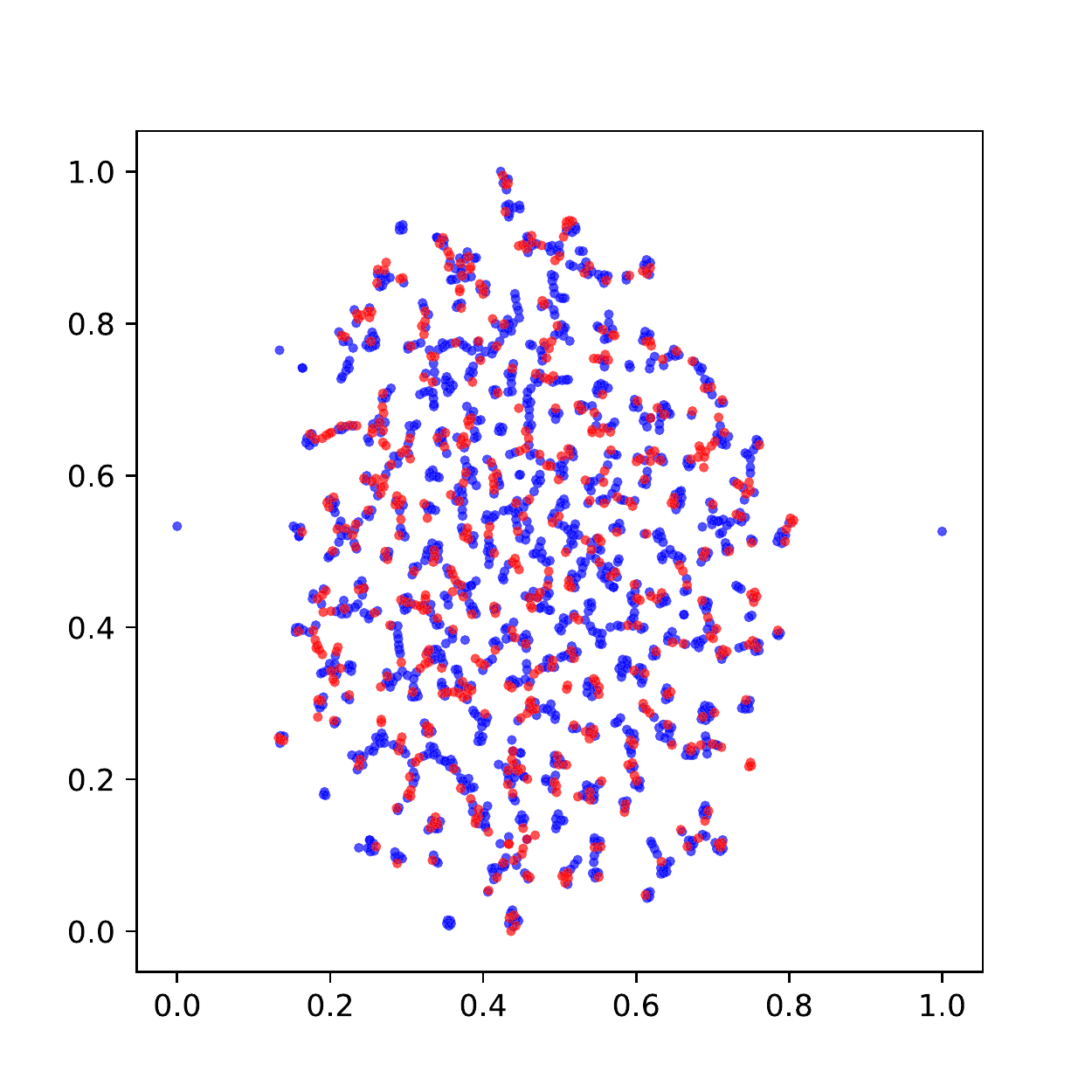}
      \caption{XBully}
    \end{subfigure}%
    \begin{subfigure}{.25\textwidth}
    \centering
      \includegraphics[width=.9\linewidth]{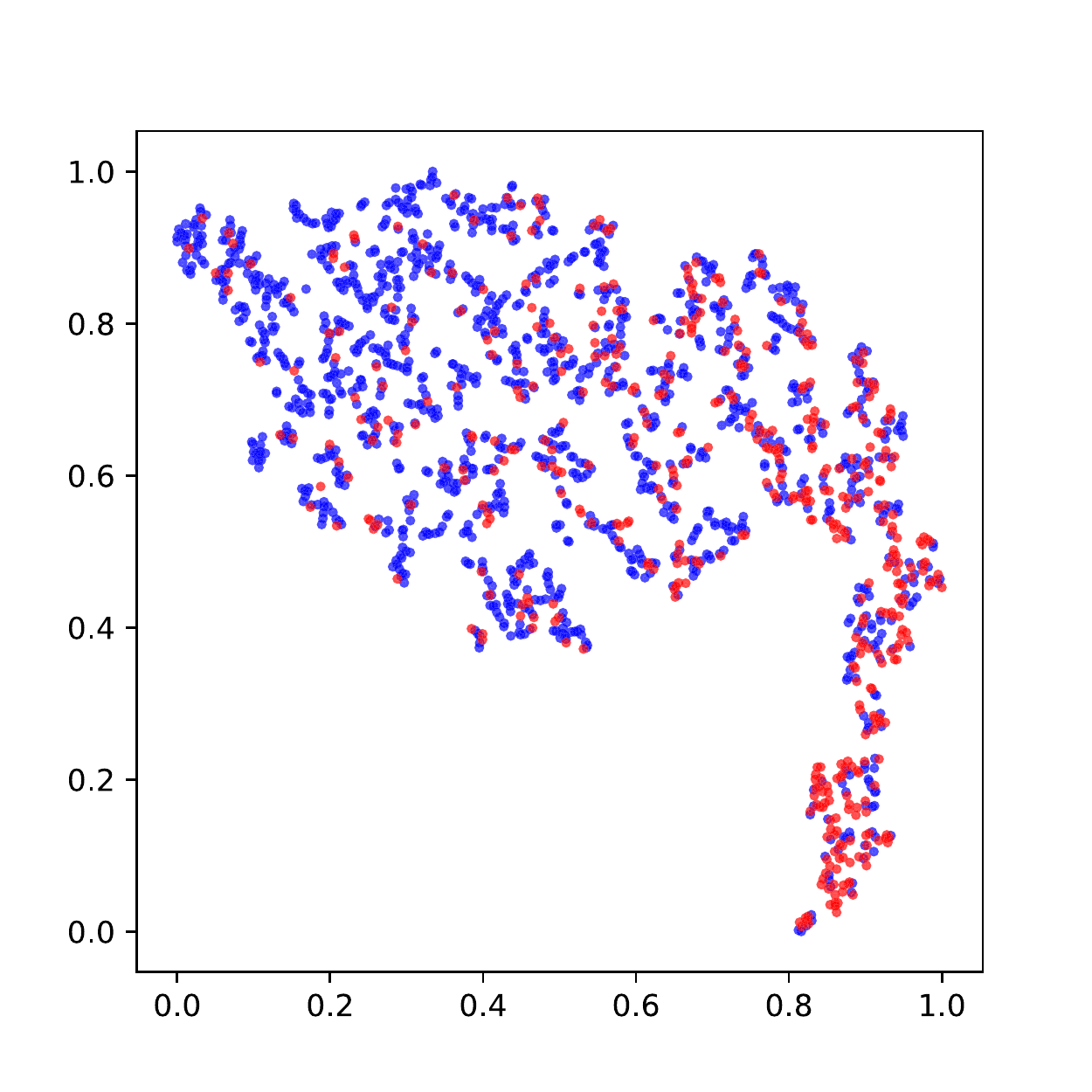}
      \caption{HAE}
    \end{subfigure}%
        \begin{subfigure}{.25\textwidth}
    \centering
      \includegraphics[width=.9\linewidth]{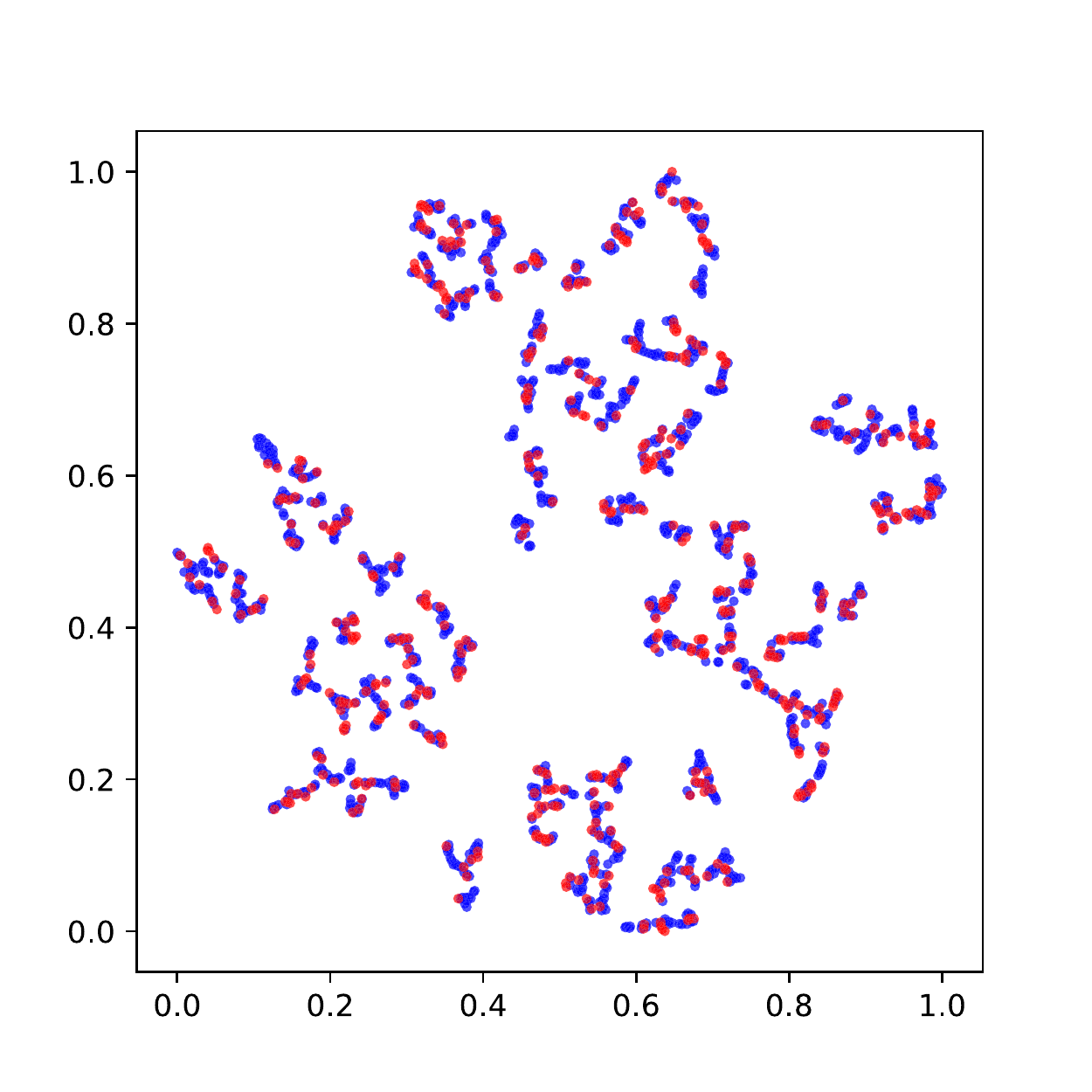}
      \caption{DCN}
    \end{subfigure}%
     \begin{subfigure}{.25\textwidth}
    \centering
      \includegraphics[width=.9\linewidth]{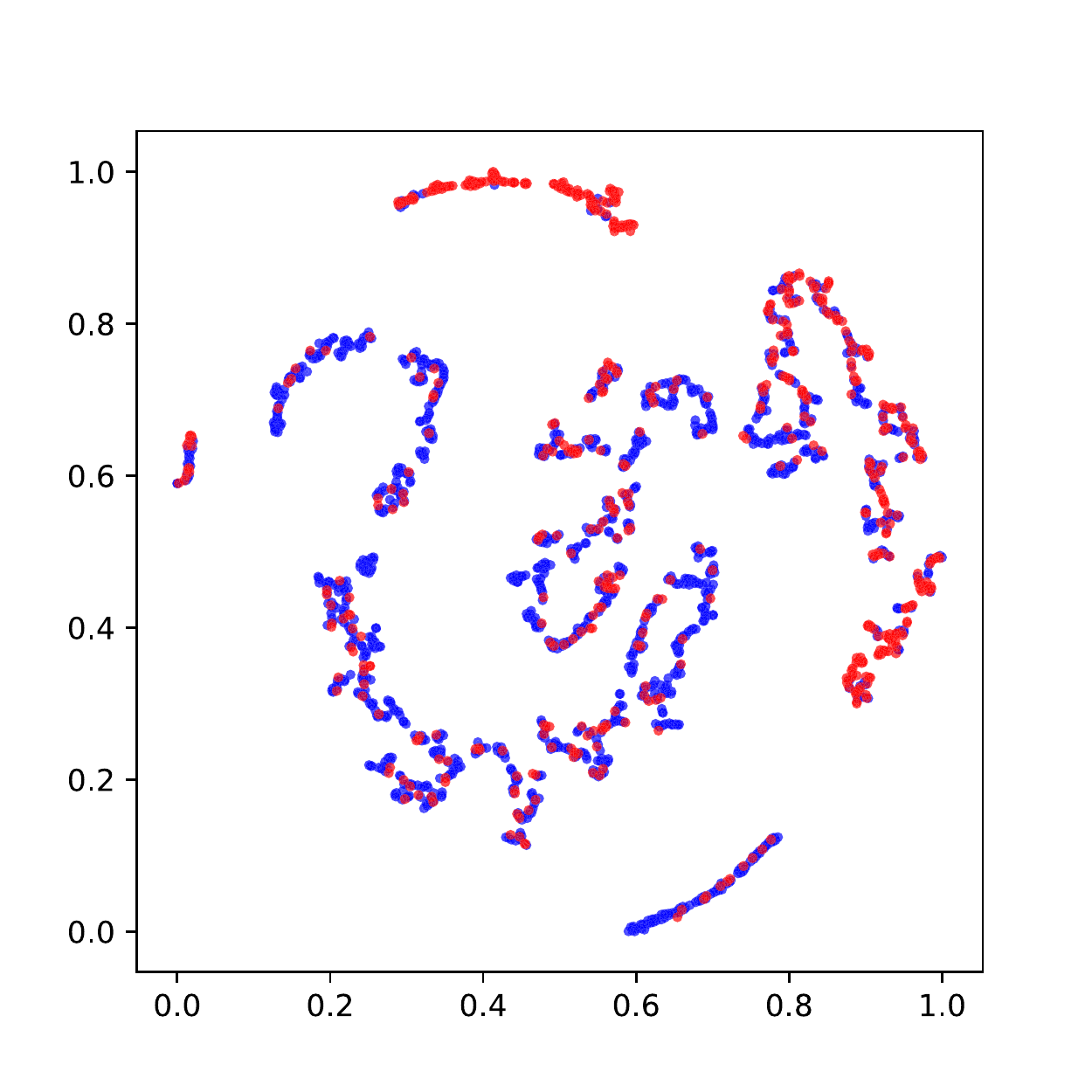}
      \caption{DAGMM}
    \end{subfigure}
    \begin{subfigure}{.25\textwidth}
    \centering
      \includegraphics[width=.9\linewidth]{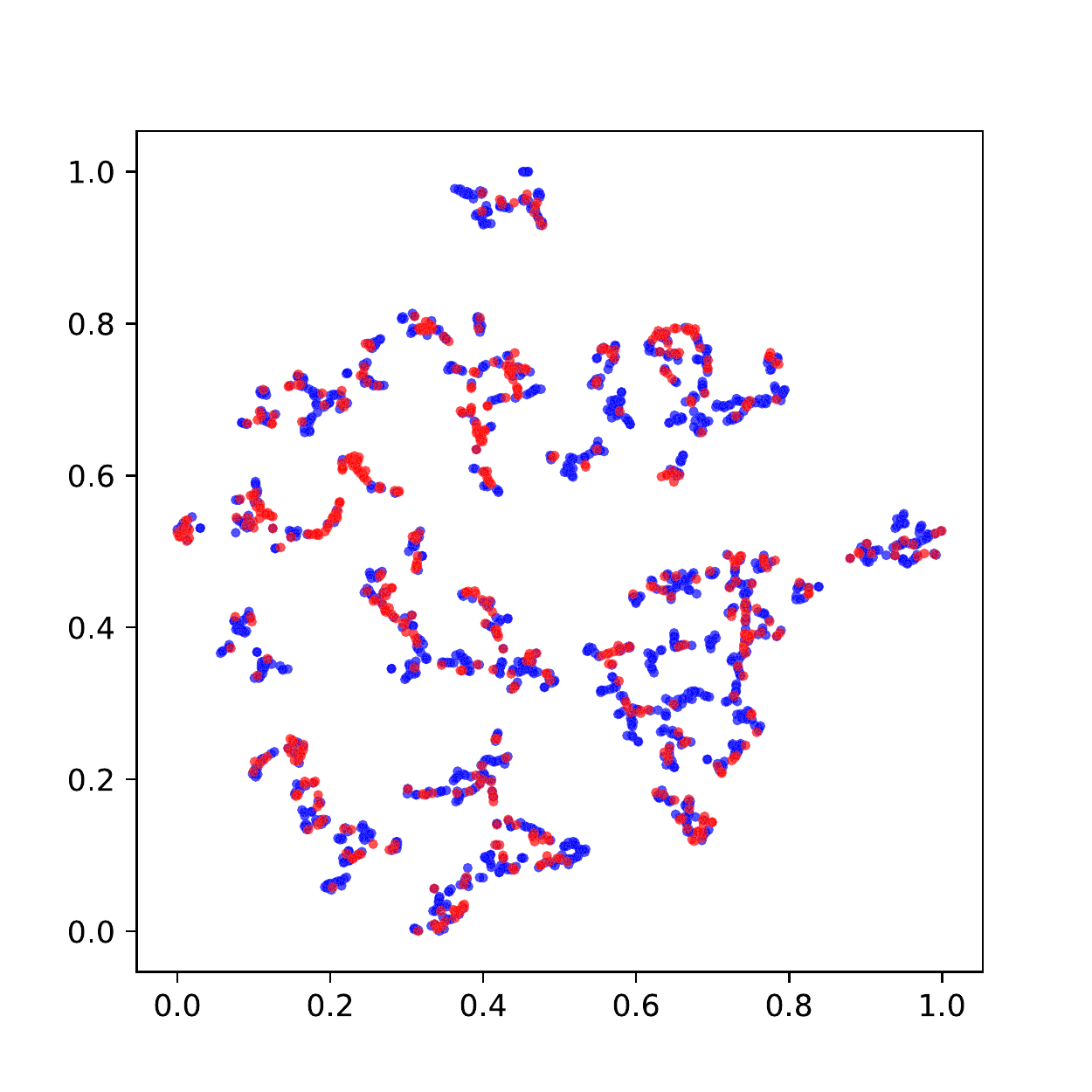}
      \caption{UCDXtext}
    \end{subfigure}%
    \begin{subfigure}{.25\textwidth}
    \centering
      \includegraphics[width=.9\linewidth]{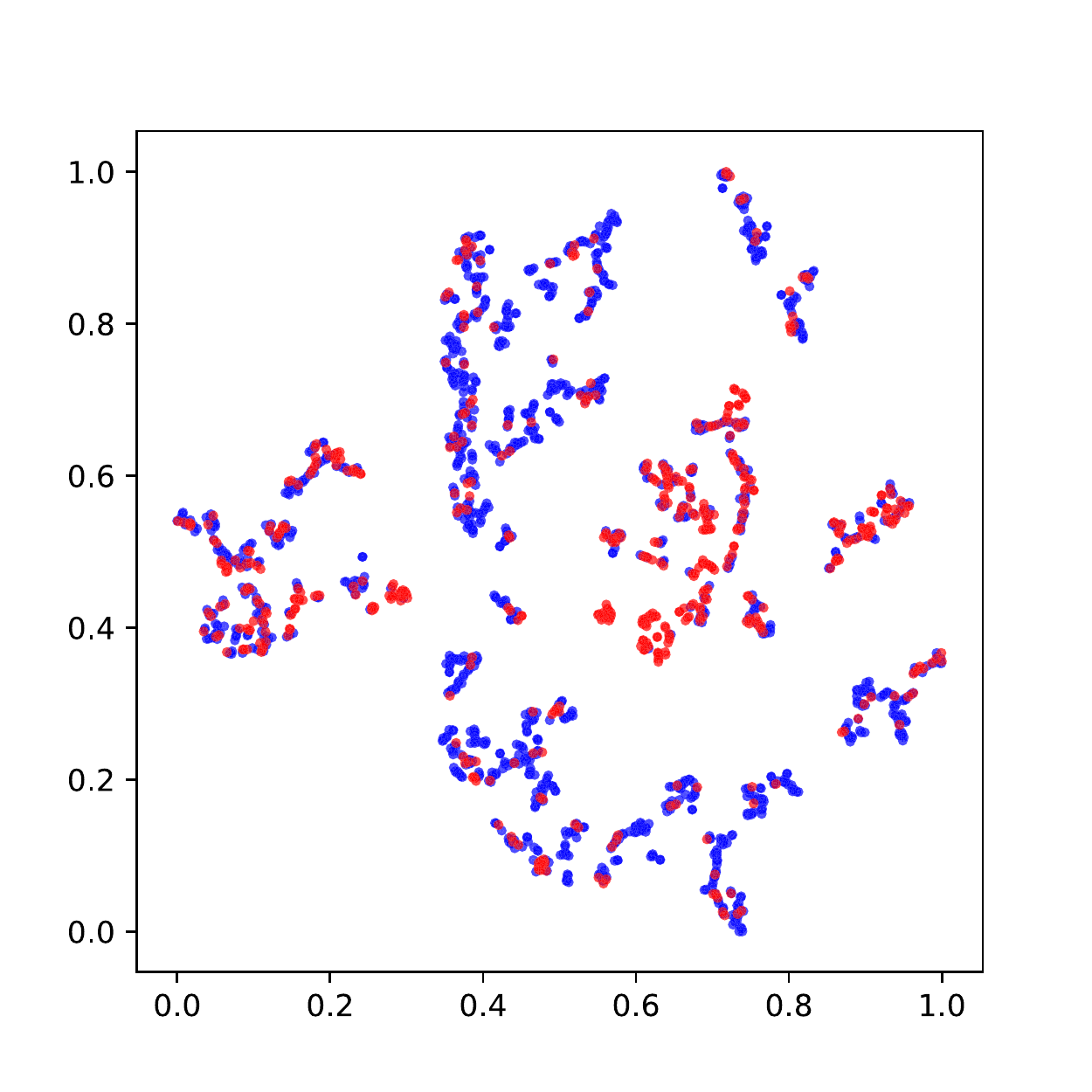}
      \caption{UCDXtime}
    \end{subfigure}%
    \begin{subfigure}{.25\textwidth}
    \centering
      \includegraphics[width=.9\linewidth]{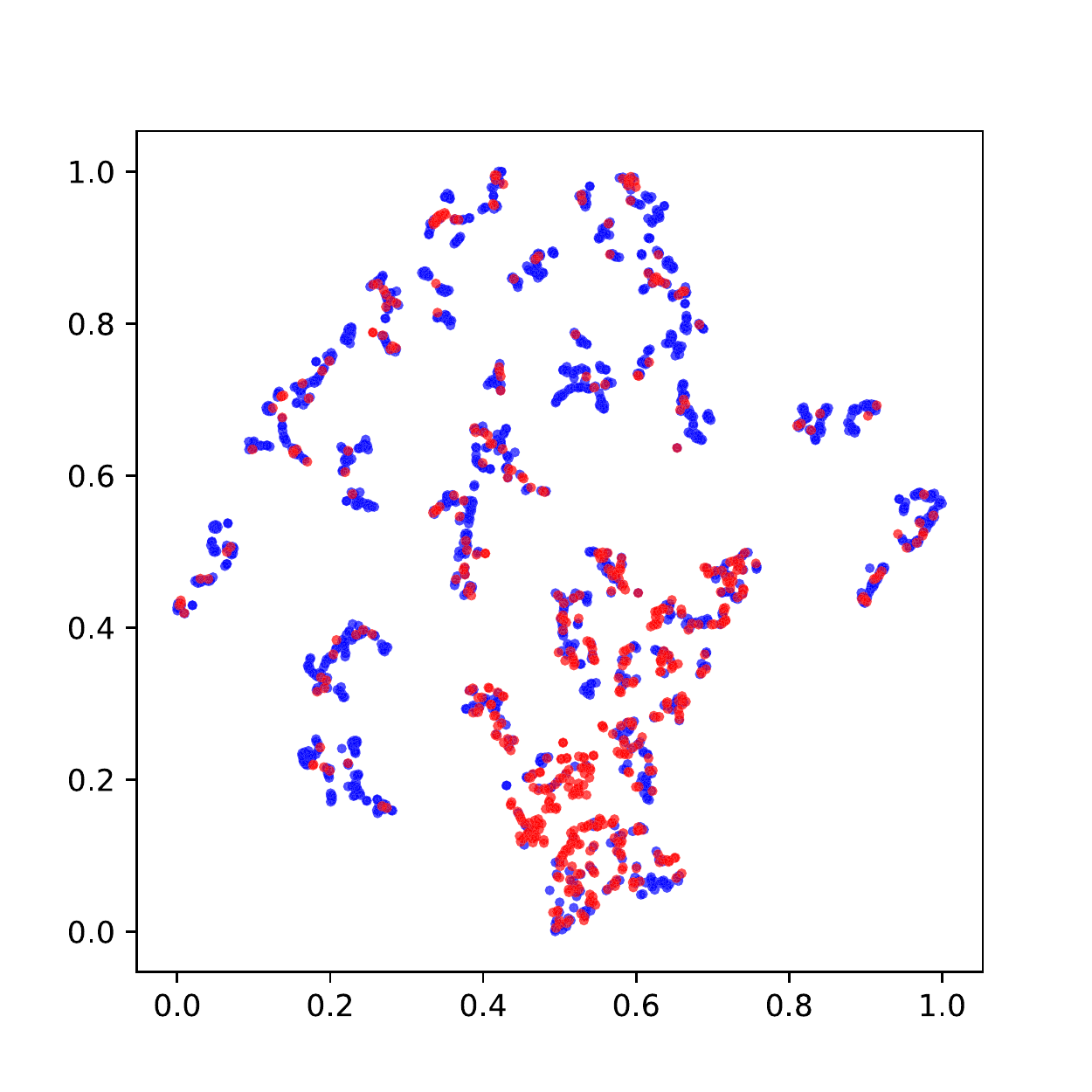}
      \caption{UCDXgraph}
    \end{subfigure}%
    \begin{subfigure}{.25\textwidth}
    \centering
      \includegraphics[width=.9\linewidth]{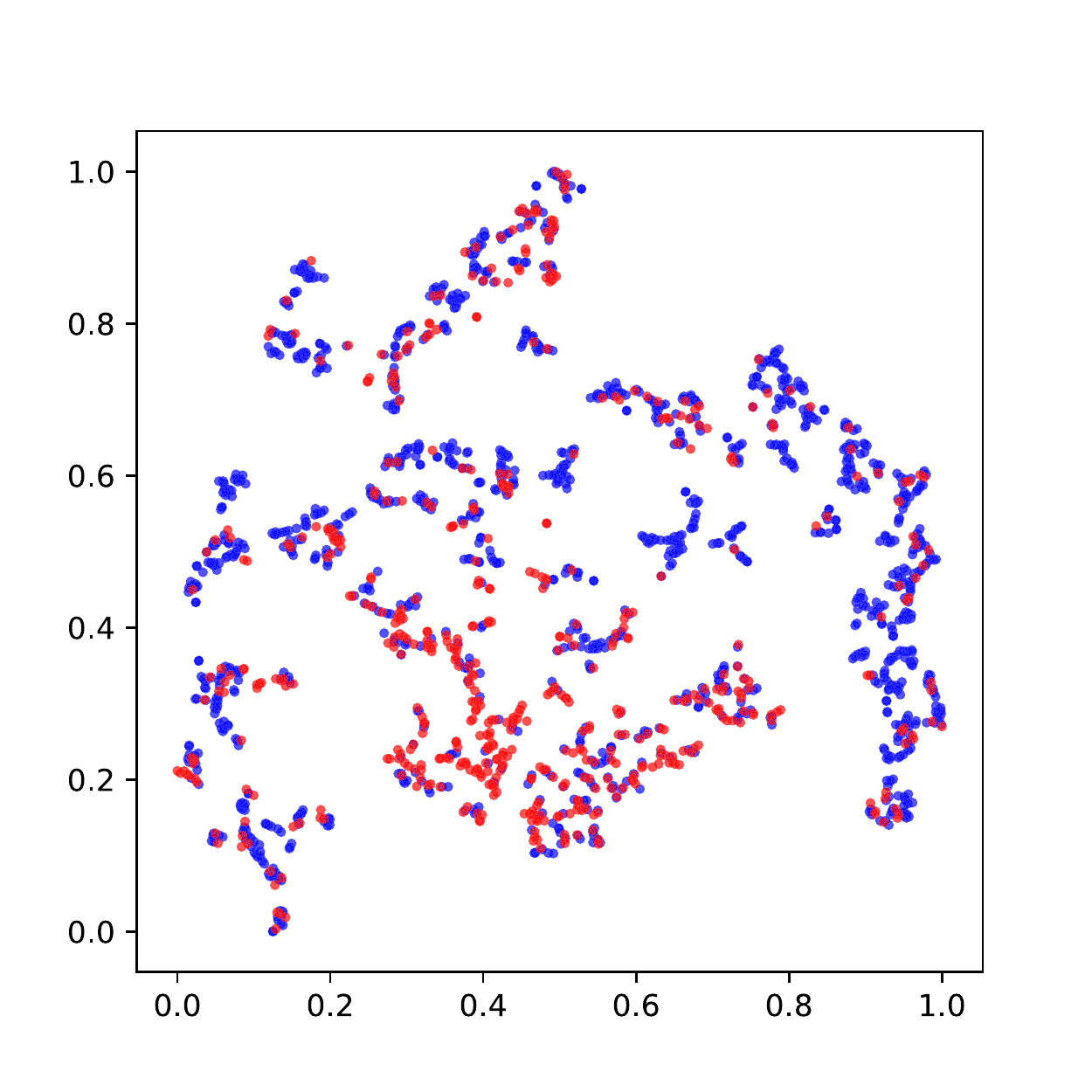}
      \caption{UCD}
    \end{subfigure}
    \caption{t-SNE visualizations of the low dimensional representations using the \textit{Instagram} dataset. The red dots denote instances of the bullying class and the blue points instances of the non-bullying class. Best viewed in colors.}
    \label{fig tsne}
\end{figure*}
\subsection{Qualitative Analysis}
We further investigate the qualities of the learned multi-modal representations using t-SNE visualizations in Fig. \ref{fig tsne}. Taking \textit{Instagram} as an example, we make the following observations:
\begin{itemize}[leftmargin=*]
    \item As shown in Fig. \ref{fig tsne}(h), UCD better separates the bullying and non-bullying samples in the latent space. The results of most of the other models, particularly XBully, HAE, DCN, and UCDXtext, yield more overlapped clusters. 
    \item From the results of DAGMM and UCD, we observe that models with GMM can learn discriminative representations, which is evident by the greater separation between bullying and non-bullying clusters). The overall performance of UCD is better than DAGMM, indicating that UCD benefits from the joint optimization of cyberbullying detection and time interval prediction.
    \item Both UCD and DAGMM outperform DCN. With a pre-trained auto-encoder, DCN can get easily stuck in a local optimum for achieving lower reconstruction error and could be suboptimal for the subsequent density estimation tasks \cite{zong2018deep}. A joint optimization of representation learning, bullying-energy estimation, and time interval prediction can help avoid these local optimal cases and achieve better learning performance.
    \item In contrast to other baseline methods, such as XBully and DCN, HAE in Fig. \ref{fig tsne}(b) generates large regions that are primarily populated by either bullying or non-bullying samples. This confirms that modeling the hierarchical structure of a session has an important impact in cyberbullying detection. 
    \item UCDXtime produces two main bullying clusters (two red clusters), UCDXgraph generates similar results to UCD, and UCDXtext fails to learn discriminative representations, evidenced by the overlap between the bullying and non-bullying clusters.
\end{itemize}
% To summarize, the proposed UCD framework -- which jointly optimizes representation learning, bullying energy estimation, and temporal dynamics prediction -- enables the representation learning and multi-task learning components to mutually boost their performance during the training process, resulting in the highest Recall, F1, and AUROC scores of the unsupervised models.
\subsection{Parameter Analysis}
\begin{figure}
\captionsetup{justification=centering}
\centering
    \begin{subfigure}{.5\columnwidth}
    \centering
      \includegraphics[width=\linewidth]{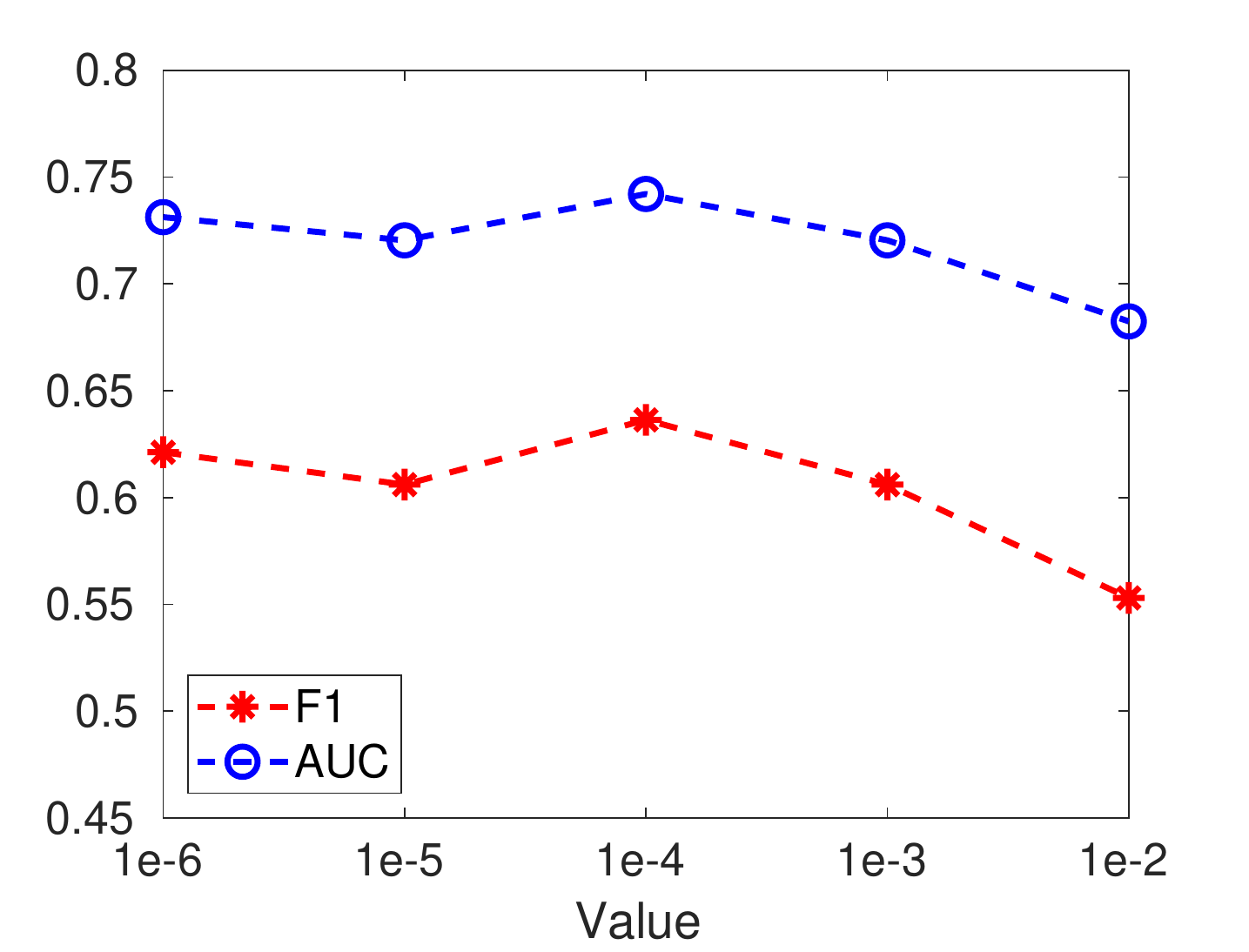}
      \caption{Effect of $\lambda_1$}
    \end{subfigure}%
    \begin{subfigure}{.5\columnwidth}
    \centering
      \includegraphics[width=\linewidth]{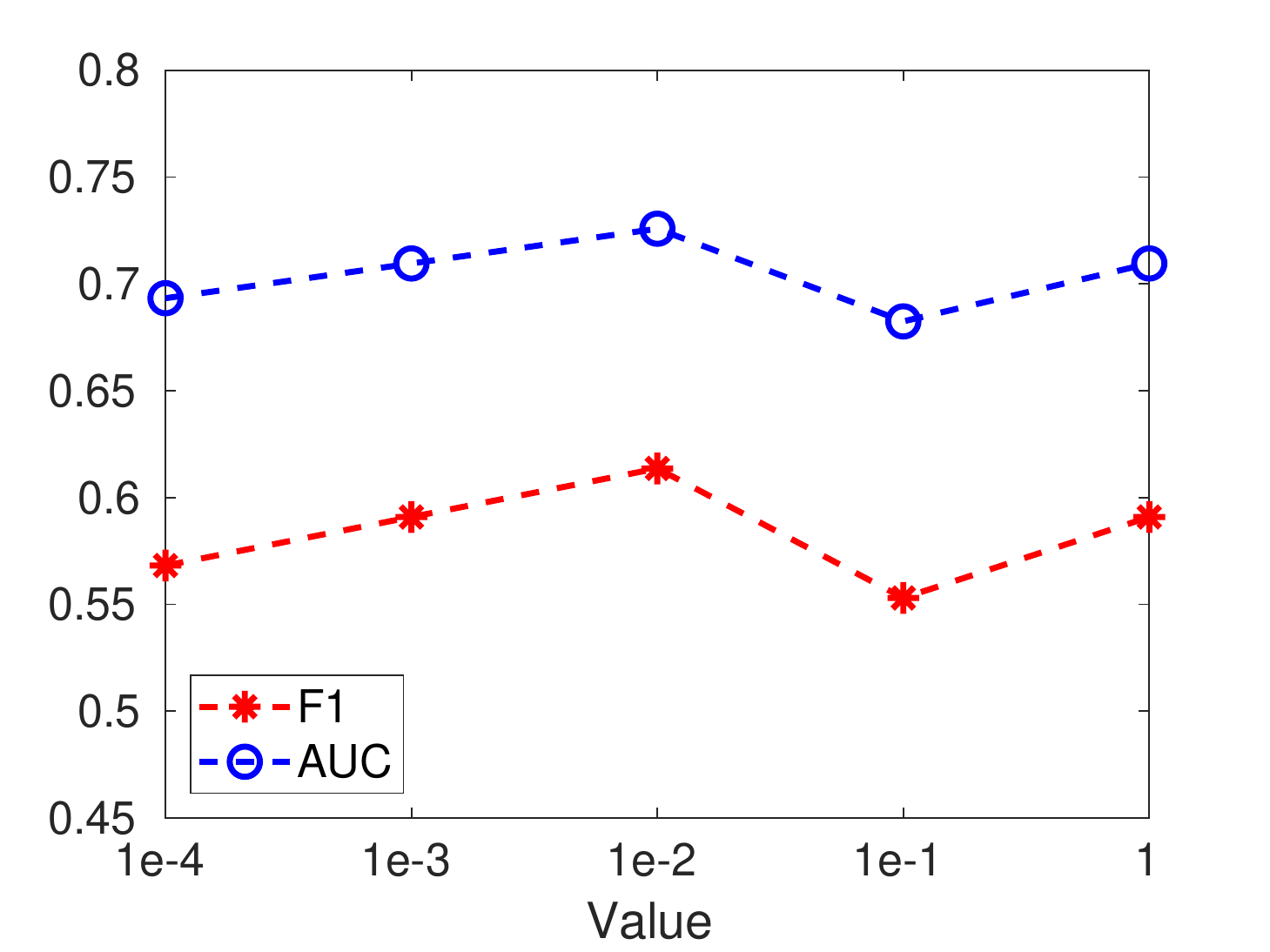}
      \caption{Effect of $\lambda_2$}
    \end{subfigure}
    \begin{subfigure}{.5\columnwidth}
    \centering
      \includegraphics[width=\linewidth]{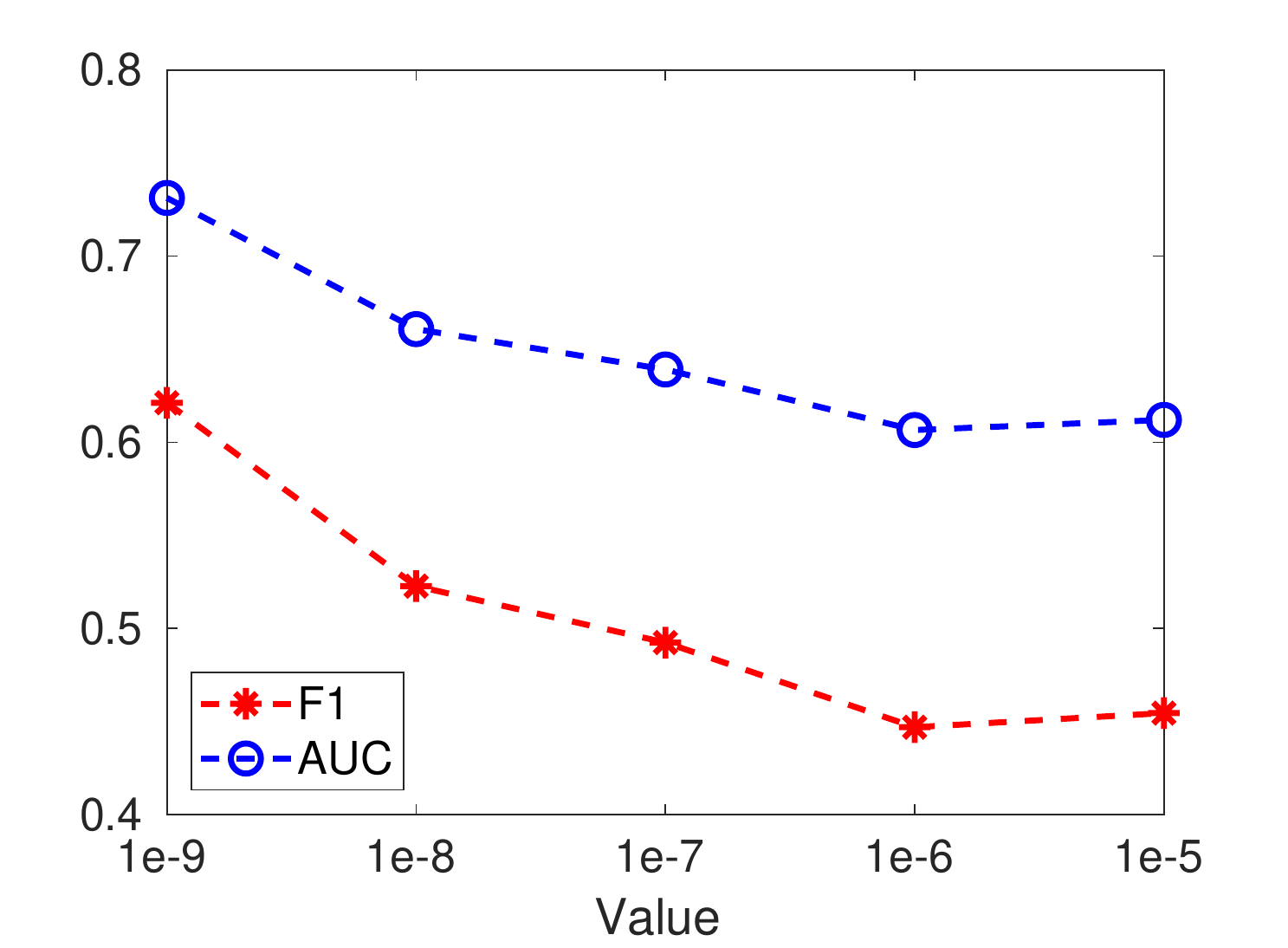}
      \caption{Effect of $\lambda_3$}
    \end{subfigure}%
    \begin{subfigure}{.5\columnwidth}
    \centering
      \includegraphics[width=\linewidth]{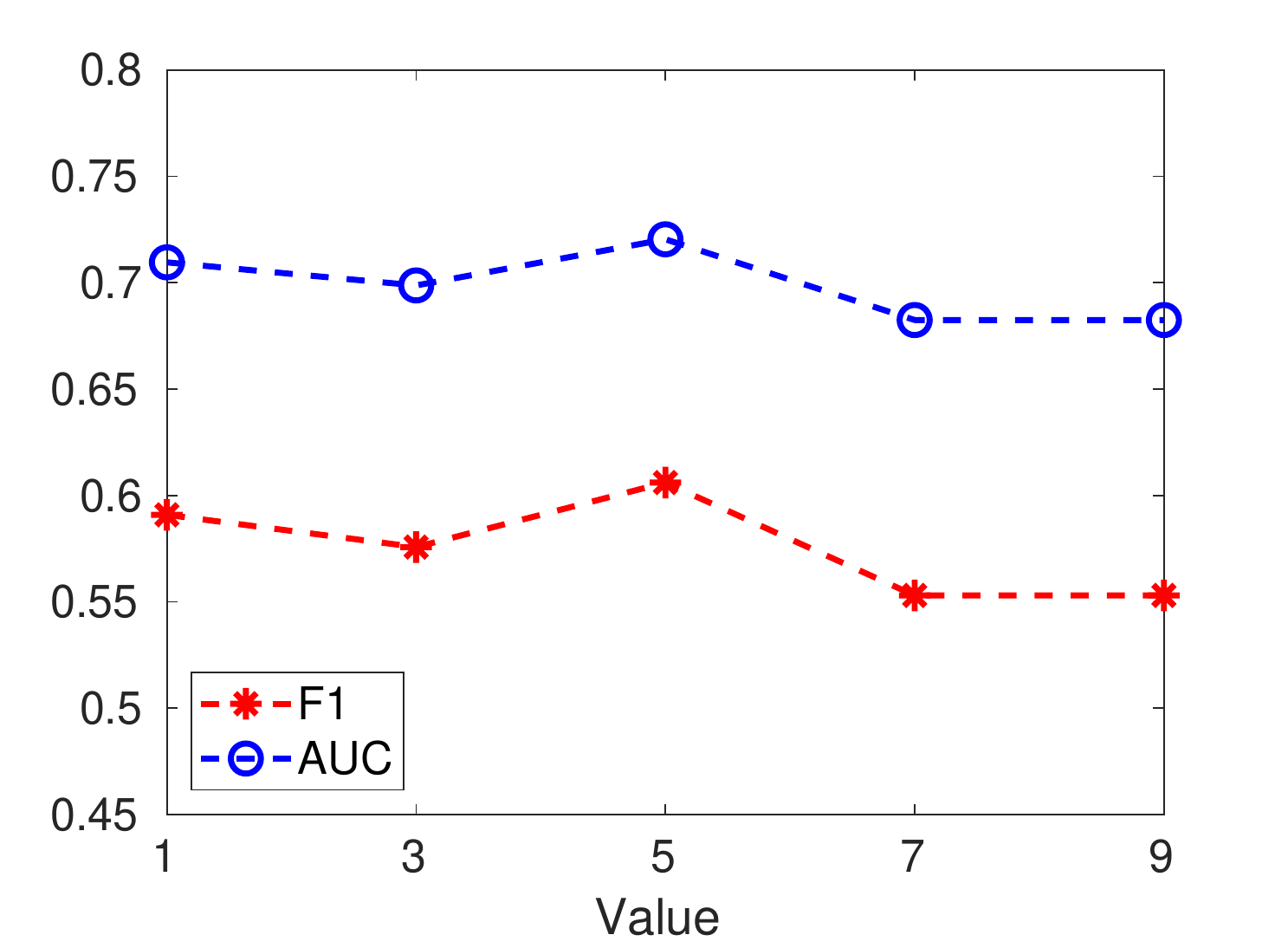}
      \caption{Effect of $K$}
    \end{subfigure}
    \caption{Parameter study w.r.t the AUROC and F1 scores.}
    \label{fig senAna}
\end{figure}

\begin{figure}
\captionsetup{justification=centering}
\centering
    \begin{subfigure}{.5\columnwidth}
    \centering
      \includegraphics[width=.85\linewidth]{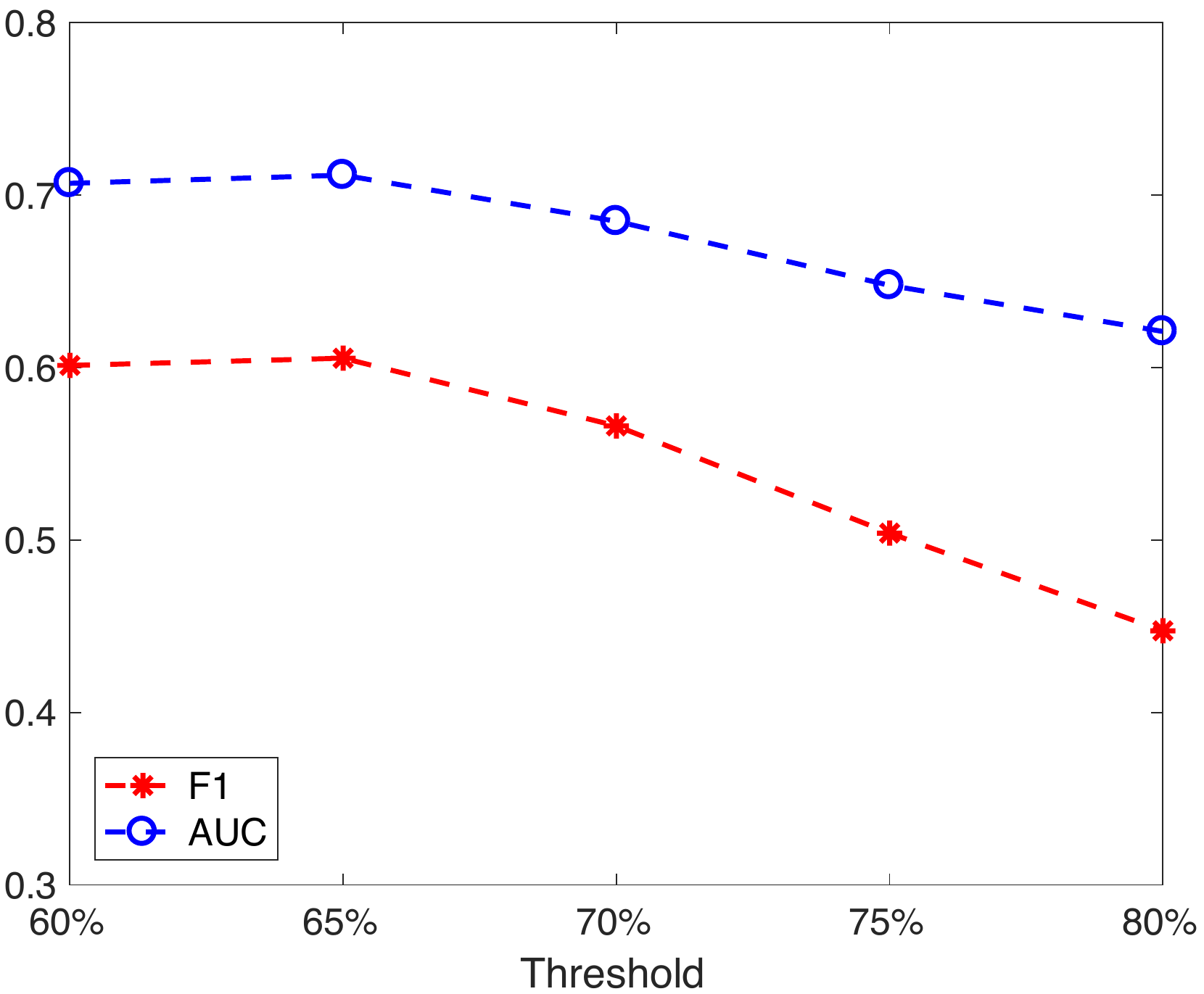}
      \caption{\textit{Instagram}}
    \end{subfigure}%
    \begin{subfigure}{.5\columnwidth}
    \centering
      \includegraphics[width=.85\linewidth]{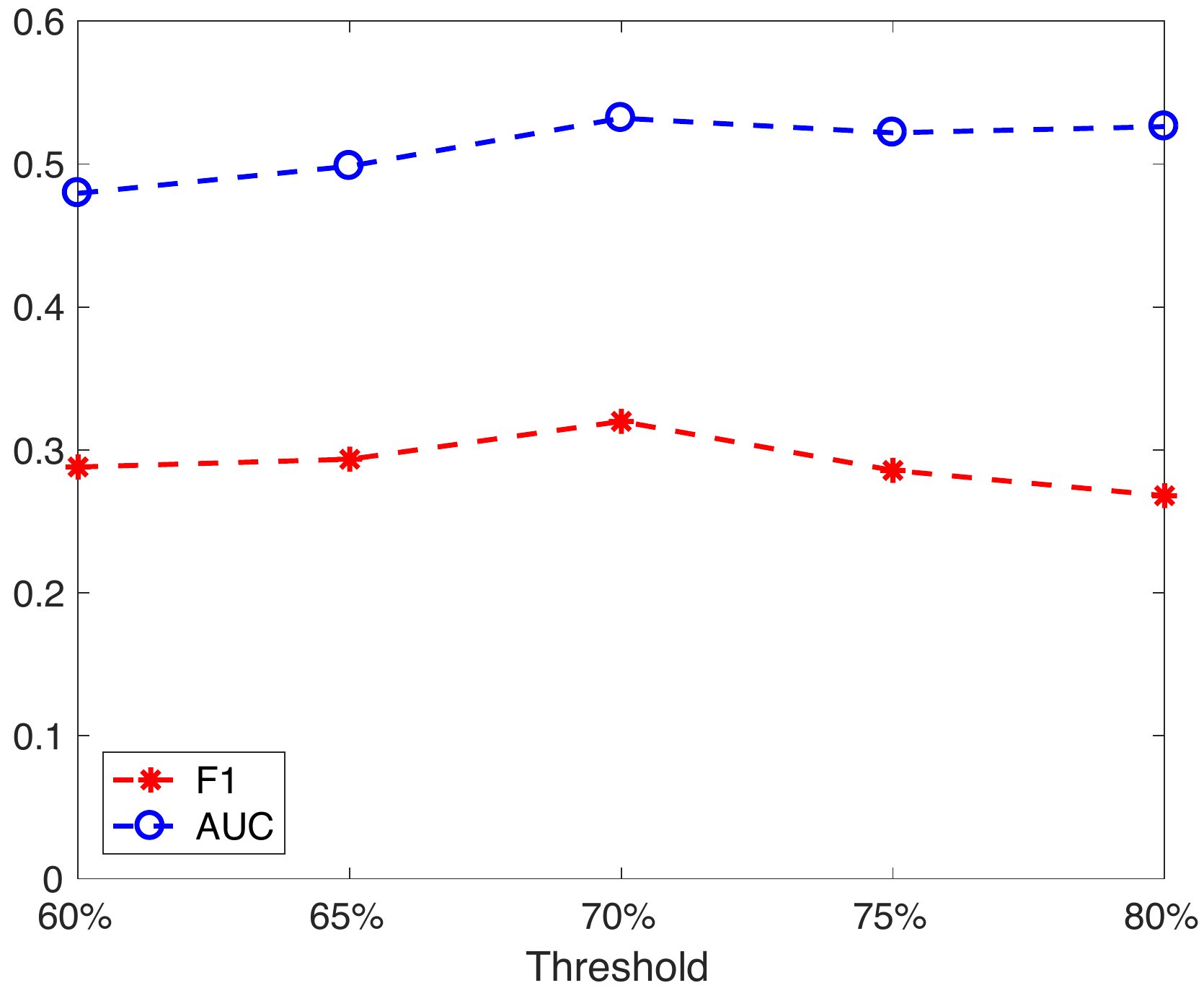}
      \caption{\textit{Vine}}
    \end{subfigure}
\caption{Effects of $\tau$ on AUROC and F1 scores.}
    \label{threshold}
\end{figure}

The UCD model has five core parameters ($\lambda_1$, $\lambda_2$, $\lambda_3$, $K$, $\tau$) for balancing the weights of bullying-energy estimation loss, reconstruction error, regularization of the covariance matrices, the number of mixtures in GMM, and the energy threshold, respectively. Here, we further divide the training data into training (80\%) and validation (20\%) sets. To investigate the effects of the first four parameters, we run experiments on the \textit{Instagram} dataset varying one parameter at a time and evaluate how it affects the overall performance. We show the sensitivity analysis w.r.t. AUROC and F1 scores in Fig. \ref{fig senAna}. We observe that large $\lambda_1$ that overemphasizes the energy estimation loss can lead to poor performance regarding both F1 and AUROC scores. The trend of varying $K$ is similar to that of $\lambda_1$, i.e., the performance drops when the number of components in GMM becomes too large. The best performance is obtained when $\lambda_1$ is set to $1\mathrm{e}{-}4$ and $K$ is set to 5. In contrast, the performance of varying $\lambda_2$ displays an ascending trend in a certain range as shown in Fig. \ref{fig senAna}(b). The UCD model with a slightly large $\lambda_2$ controlling the importance of GAE is more likely to obtain better results. Unsurprisingly, when the covariance matrices in GMM are given too much penalization, i.e., a large $\lambda_3$, the F1 and AUROC scores decrease significantly, as shown in Fig. \ref{fig senAna}(c). The last parameter $\tau$ represents the threshold for identifying bullying instances. Given that UCD largely relies on $\tau$ for cyberbullying detection, we use both \textit{Instagram} and \textit{Vine} datasets to examine its influence. The results are presented in Fig. \ref{threshold}. It shows that UCD is more robust to $\tau$ for \textit{Vine}, whereas its performance slightly decreases for \textit{Instagram} as $\tau$ increases. In practice, $\lambda_3$ should be set to a small value, and a proper value for parameter $\tau$ should be experimentally identified. In general, UCD is robust to most of the model parameters, and consequently can be tuned for various real-world applications.
\subsection{Case Study}
\begin{figure}
\captionsetup{justification=centering}
\centering
    \begin{subfigure}{\columnwidth}
    \centering
      \includegraphics[width=\linewidth]{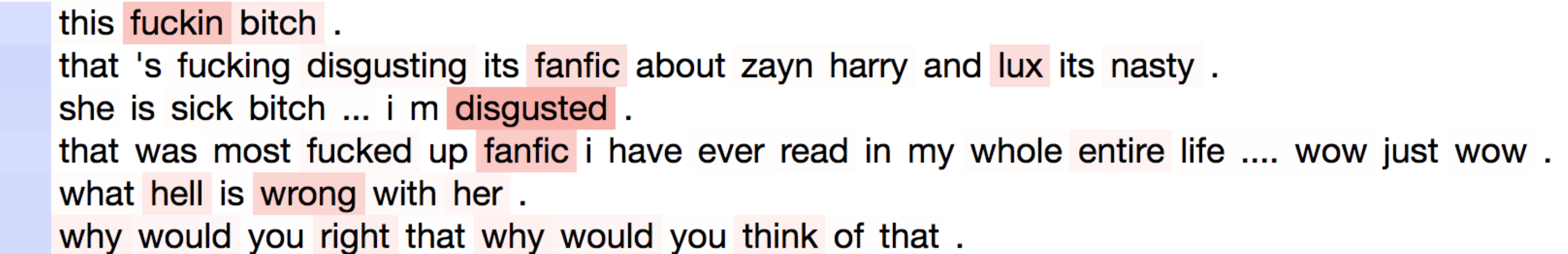}
      \caption{\textit{Predicted as bullying session.}}
      \label{bully_att}
    \end{subfigure}
    \begin{subfigure}{\columnwidth}
    \centering
      \includegraphics[width=\linewidth]{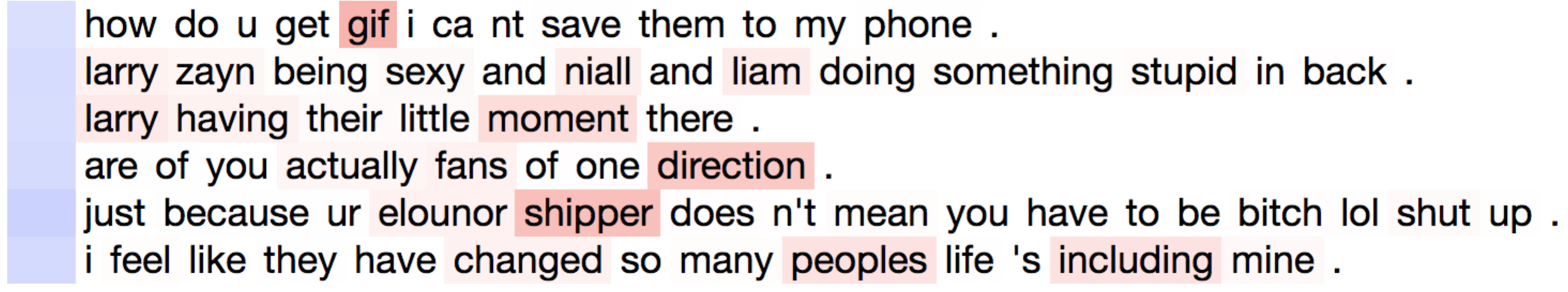}
      \caption{\textit{Predicted as non-bullying session.}}
    \end{subfigure}
\caption{Case study using the \textit{Instagram} dataset.}
    \label{attention}
\end{figure}

In this subsection, we present two Instagram sessions, one detected as bullying and one detected as non-bullying by UCD. We visualize each with the hierarchical attention information to validate UCD's capability of selecting informative comments and words in a session. The results can be seen in Fig. \ref{attention}. Every line in each sub-figure is a comment. Shades of blue denote comment weights and shades of red denote word weights. Because both sessions have many comments, only a portion of the content is shown here. Fig. \ref{attention}(a) shows that UCD can select the words that are more strongly associated with bullying, such as \textit{f*ckin, b*tch, disgusted} and \textit{hell}. In Fig. \ref{attention}(b), we observe that UCD can also deal with complex cross-comment context. For example, although the session might appear to be a bullying session when looking only at the second comment from the bottom, UCD assigns the session to the non-bullying cluster because it also considers the context of that comment.
\section{Discussion}
In this section, we elaborate on the reasons behind the performance of UCD, its research impact, and practical considerations. UCD benefits from the following design mechanisms: 
\begin{itemize}[leftmargin=*]
    \item \textit{Multi-modal features.} UCD actively leverages multi-modal data including text, user information, social network information, and social content. UCD also benefits from deep learning mechanisms specifically designed for each modality, e.g., HAN models the sequence of comments and the hierarchy of a session. Previous work \cite{cheng2019xbully} reported the benefits of using multi-modal data to contribute complementary application domain insights and enable better learning performance.
    \item \textit{Complementary temporal analysis.} In addition to multi-modal representation learning, UCD simultaneously estimates the energy level associated with bullying instances and predicts the time-interval between comments to refine the session representations. Temporal modeling adds nuance to the representation learning network that otherwise would not consider comment evolution \cite{soni2018time,cheng2019hierarchical}.
    \item \textit{Joint optimization.} A key property that differentiates UCD from other approaches is that it jointly optimizes the parameters for representation learning, temporal modeling, and bullying-energy estimation. This approach prevents the drawbacks of decoupled training.
\end{itemize}
%OLD text: In addition to multi-modal representation learning, UCD simultaneously estimates sample energy of being a bullying instance and predicts ...

As one of the first attempts to detect cyberbullying in an unsupervised manner, UCD explores the use of deep learning algorithms and shows they can achieve relatively high performance levels. The development of UCD has relevant research and practical impact. UCD addresses key limitations of supervised models: (1) cyberbullying labeled data could be either unavailable or insufficient for training a good supervised classifier, (2) data labeling is often time-consuming and labor-intensive, and (3) the guidelines used for assigning cyberbullying labels in a current session cannot always be generalized to future sessions due to the dynamic nature of language and social networks. We hope this work will motivate further research efforts in unsupervised cyberbullying detection. Regarding the practical use of UCD, it could be integrated into third-party anti-bullying apps, such as Bark\footnote{https://www.bark.us} and BullyBlocker \cite{silva2018bullyblocker}, or as a component of automated mediation tools \cite{qian2019benchmark}.
%OLD Text: As one of the first attempts to detect cyberbullying in an unsupervised manner, UCD explores the use of deep learning algorithms and shows they can achieve relatively high performance levels. The development of UCD has relevant research and practical impact. UCD addresses two key limitations of supervised models--that (1) given data labeling is often time-consuming and labor-intensive, UCD becomes an effective alternative when labeled data is either unavailable or insufficient for training a good supervised classifier; and (2) the current guidelines for labeling cyberbullying cannot be generalized to future instances due to the dynamic nature of language and social networks. This study warrants further research efforts in unsupervised cyberbullying detection. Regarding the practical use of UCD, it could be easily integrated into third-party anti-bullying apps, such as Bark\footnote{https://www.bark.us} and BullyBlocker \cite{silva2018bullyblocker}, or as a component of automated mediation tools \cite{qian2019benchmark}. 

\section{Conclusions and Future Work}
Existing efforts towards detecting cyberbullying have focused primarily on supervised methods that require large amounts of time and labor to annotate datasets. To address this limitation, we propose an unsupervised cyberbullying detection framework, called UCD, which consists of two major components: a \textit{representation learning} network that encodes multi-modal session representations and a \textit{multi-task learning} network that simultaneously estimates bullying likelihood and models the temporal dynamics of arriving comments. The joint parameter optimization in UCD yields better performance. Our experimental results on two real-world datasets corroborate the effectiveness of UCD.
%OLD Text: Existing efforts towards detecting cyberbullying have focused primarily on supervised methods that require large amounts of time and labor to annotate datasets, and use annotations that may not be valid in the future due to the dynamic aspects of language and social network. To address these limitations, we propose an unsupervised cyberbullying detection framework (UCD) that consists of two major components: a \textit{representation learning network} that encodes multi-modal session representations and a \textit{multi-task learning network} that simultaneously estimates sample bullying-energy and models evolving dynamics of comments. The joint parameter optimization in UCD enables its components to more effectively contribute to better overall performance levels. Extensive experimental results on two real-world datasets corroborate UCD's effectiveness.

The presented findings elucidate multiple paths for future work, including a more detailed analysis of the temporal characteristics of cyberbullying behaviors in social media and the study of user-session networks (where users and sessions are the nodes) to directly explore the connection between users and associated posts.

\section*{Acknowledgements}
This work was in part supported by the National Science Foundation (NSF) Grants 1719722 and 1614576.
\bibliographystyle{ACM-Reference-Format}
\bibliography{sample-base}

\end{document}